\documentclass[twocolumn,useAMS,usenatbib]{mn2e}
\usepackage{graphicx}
\usepackage{bm}

\newcommand{\rmd}{{\rm d}}
\newcommand{\rme}{{\rm e}}

\newcommand{\threej}[6]{\left(\begin{array}{ccc}
  #1 & #2 & #3 \\ #4 & #5 & #6 \end{array}\right)}
\newcommand{\sixj}[6]{\left\{\begin{array}{ccc}
  #1 & #2 & #3 \\ #4 & #5 & #6 \end{array}\right\}}

\newcommand{\nh}{n_{\rm H}}
\newcommand{\xhi}{x_{\rm HI}}

\newcommand{\hi}{H$\,${\sc i}}

\newcommand{\mlya}{{{\rm Ly}\alpha}}
\newcommand{\lya}{Ly$\alpha$}
\newcommand{\twos}{$2s$}
\newcommand{\wuf}{Wouthuysen-Field}

\title[Wouthuysen-Field effect]{Wouthuysen-Field coupling strength
  and application to high-redshift 21~cm radiation}

\author[Hirata]
 {Christopher M. Hirata\thanks{Electronic address:
    {\tt chirata@sns.ias.edu}}
\\Institute for Advanced Study, Einstein Drive,
      Princeton, NJ 08540, USA
}

\date{22 November 2005}

\begin{document}
\maketitle

\begin{abstract}
The first ultraviolet sources in the universe are expected to have coupled 
the \hi\ spin temperature to the gas kinetic temperature via scattering in 
the \lya\ resonance (the ``\wuf\ effect'').  By establishing an \hi\ spin 
temperature different from the temperature of the cosmic microwave 
background, the \wuf\ effect should allow observations of \hi\ during the 
reionization epoch in the redshifted 21~cm hyperfine line.  This paper 
investigates four mechanisms that can affect the strength of the \wuf\ 
effect that were not previously considered: (1) Photons redshifting into 
the \hi\ Lyman resonances may excite an H atom and result in a radiative 
cascade terminating in two-photon $2s_{1/2}\rightarrow 1s_{1/2}$ emission, 
rather than always degrading to \lya\ as usually assumed.  (2) The fine 
structure of the \lya\ resonance alters the photon frequency distribution 
and leads to a suppression of the scattering rate.  (3) The spin-flip 
scatterings change the frequency of the photon and cause the photon 
spectrum to relax not to the kinetic temperature of the gas but to a 
temperature between the kinetic and spin temperatures, effectively 
reducing the strength of the \wuf\ coupling.  (4) Near line centre, a 
photon can change its frequency by several times the line width in a 
single scattering event, thus potentially invalidating the usual 
calculation of the \lya\ spectral distortion based on the diffusion 
approximation. It is shown that (1) suppresses the \wuf\ coupling strength 
by a factor of up to $\sim 2$, while 
(2) and (3) are important only at low kinetic temperatures.
Effect (4) has a $\le 3$ per cent effect for 
kinetic temperatures $T_k\ge 2\,$K.  In 
particular if the pre-reionization intergalactic medium was efficiently 
heated by X-rays, only effect (1) is important.  Fitting formulae for the 
\wuf\ coupling strength are provided for the range of $T_k\ge 2\,$K and 
Gunn-Peterson optical depth $10^5<\tau_{GP}<10^7$ so that 
all of these effects can be easily incorporated into 21~cm codes.
\end{abstract}


\section{Introduction}

The cosmic reionization is one of the unexplored frontiers of 
astrophysics.  Currently we have only a few limited observational 
constraints on the nature of the intergalactic medium (IGM) during this 
era and the objects that must have formed during it.  The major 
constraints on reionization currently come from the \hi\ \lya\ absorption 
at a wavelength of $\lambda_\mlya=1216\,$\AA\ and from the polarization of 
the cosmic microwave background (CMB).  In particular, observations of 
complete \lya\ absorption at $z\sim 6$ in quasar spectra have pinpointed 
this epoch as the end of reionization \citep{2001AJ....122.2850B, 
2002AJ....123.1247F}, whereas the CMB polarization data from the {\slshape 
Wilkinson Microwave Anisotropy Probe} ({\slshape WMAP}) suggest a 
significant ionization at higher redshifts, e.g. for instantaneous 
reionization {\slshape WMAP} finds reionization at $z=20^{+10}_{-9}$ 
\citep{2003ApJS..148....1B, 2003ApJS..148..161K}.

While the \lya\ and {\slshape WMAP} polarization data are currently our 
best source of information about the early ionization history of the IGM 
and the ionizing sources responsible for reionization, these techniques 
leave several fundamental questions unanswered.  The \lya\ absorption 
saturates at relatively low neutral fraction $\xhi\ll 1$ and cannot probe 
the bulk of the reionization epoch.  The CMB polarization does probe the 
bulk of the reionization epoch, but on the large angular scales of 
interest, cosmic variance limits the precision with which information can 
be extracted \citep{2003PhRvD..68b3001H} and polarized foregrounds may 
prove to be a further limitation.  The large-scale polarization also only 
probes the mean ionization of the universe, and has coarse redshift 
information.  CMB anisotropies on small scales are sensitive to patchy 
reionization, but these come with no redshift information, so their 
interpretation could be difficult \citep{2004ApJ...606...46D}.

One promising source of information about the reionization history that 
overcomes both of these problems is the hyperfine 21.1~cm line of \hi.  
For most of the reionization era, \hi\ is present in significant 
quantities.  Moreover radio interferometry may make 21~cm inhomogeneities 
observable across a range of angular scales, and because the 21~cm 
radiation is a spectral line, frequency information immediately gives the 
redshift.  Thus the 21~cm line has attracted much interest as a probe of 
the high-redshift IGM \citep{1979MNRAS.188..791H, 1997ApJ...475..429M, 
2003MNRAS.341...81I, 2003ApJ...596....1C, 2004ApJ...608..622Z, 
2004ApJ...610..117W}.  Several experiments are currently being built or 
planned to observe the high-redshift 21~cm signal, including the Primeval 
Structure Telescope \citep{2004MPLA...19.1001P}, the Low-Frequency 
Array\footnote{http://www.lofar.org/}, the Mileura Wide-field 
Array\footnote{http://web.haystack.mit.edu/arrays/MWA/index.html}, and the 
Square Kilometre Array\footnote{http://www.skatelescope.org/}.

The 21~cm line is sensitive to several properties of the IGM including its 
density, neutral fraction $\xhi$, and spin temperature $T_s$.  Before the 
first ultraviolet (UV) sources turn on, the spin temperature is determined 
by a competition between the tendency of radiative transitions to bring 
$T_s$ into equilibrium with the CMB at $T_\gamma$ and the tendency of 
atomic collisions to bring $T_s$ into equilibrium with the gas kinetic 
temperature $T_k$.  \citet{2004PhRvL..92u1301L} have shown that at 
redshifts $z\sim 30$ the radiative transitions dominate over collisions in 
regions of the universe near the mean density.  Collisions still dominate 
in the highest-density regions of the universe such as minihalos, which 
can be hotter than the CMB and thus appear in emission 
\citep{2002ApJ...572L.123I, 2005astro.ph..9651A, 2005astro.ph.10814K}.  
Therefore at these redshifts, the 21~cm signal should consist of very weak 
absorption from most of the volume, plus emission from the high-density 
regions.

However, once the first galaxies form, UV radiation is released into the 
IGM.  This radiation can Raman-scatter through the \lya\ resonances and 
convert hydrogen atoms between the two hyperfine levels $F=0$ and $F=1$.  
The photons within the \lya\ resonance region can exchange energy with 
\hi\ atoms via the Doppler shift, hence they are expected to come to 
Boltzmann equilibrium with the gas kinetic temperature, and so the Raman 
scattering should tend to bring $T_s$ into equilibrium with $T_k$.  This 
process is known as the \wuf\ effect, after \citet{1952AJ.....57R..31W} 
and \citet{field1958}; this effect, together with the CMB and collisions, 
controls the \hi\ spin temperature during reionization.  Once the \wuf\ 
effect turns on, one should observe a strong absorption signal at 
$21(1+z)\,$cm if $T_k<T_\gamma$ as expected if the IGM has expanded 
adiabatically since thermal decoupling from the CMB at $z\sim 200$, or an 
emission signal if the neutral IGM has been heated efficiently by X-rays 
\citep{1997ApJ...475..429M}.

The main purpose of this paper is to investigate in more detail the 
physics of the \wuf\ effect as applied to the high-redshift IGM.  Much 
progress in this direction has recently been made due to the work of 
\citet{2004ApJ...602....1C} and \citet{2004astro.ph.10129B}, who have 
respectively investigated the mean \wuf\ coupling rate and the 
perturbations caused by the fluctuating density of galaxies.  However, 
there are several physical effects that were neglected in these papers, 
but are are investigated here.  First, it is usually assumed that any UV 
photon emitted in the band between the Lyman edge at 912~\AA\ and \lya\ at 
1216~\AA, will redshift into a Lyman-series resonance and be degraded to 
\lya\ via a radiative cascade. However, some radiative cascades in \hi\ 
terminate in the two-photon transition from $2s_{1/2}$ to $1s_{1/2}$, and 
these produce no \lya.  It is shown that all photons emitted between 
Ly$\beta$ (1026~\AA) and Ly$\gamma$ (973~\AA), and most photons between 
Ly$\gamma$ and the Lyman edge, are ``lost'' in this way. This reduces the 
\wuf\ coupling rate since the latter is determined by the flux of \lya\ 
photons.  Here it is shown that the reduction can be as much as a factor 
of $\sim 2$ for hard source spectra.

Secondly, the photon spectrum in the vicinity of \lya\ and the associated 
spin-flip rate are considered in detail, taking into consideration the 
fine and hyperfine structure of \lya, the frequency dependence of the 
spin-flip probability (which was previously assumed to be a constant 
$\frac{4}{27}$), the $\Delta\nu=\pm 1.4\,$GHz change of frequency of 
photons during spin-flip scatterings.  Additionally the validity of 
treating the \lya\ spectral feature via the Fokker-Planck equation (i.e. 
as a diffusive process) is investigated.  These corrections are only 
important at low kinetic temperatures, since at high $T_k$ the smearing of 
the \lya\ line profile by the Doppler effect during repeated scatterings 
overwhelms the 11 GHz $2p_{1/2}$--$2p_{3/2}$ fine structure splitting and 
the even smaller hyperfine splitting.  For example they suppress the \wuf\ 
effect by $\sim 10$ per cent at $T_k=5\,$K and $\sim 1$ per cent at 
$T_k=50\,$K.  \citet{2004ApJ...602....1C} argue that X-rays from 
supernovae or X-ray binaries are likely to have heated the IGM to high 
temperatures $T_k\gg T_\gamma$ well before the end of reionization; if 
this did indeed happen, then the fine and hyperfine structure effects 
considered here are completely negligible.

One could ask whether it is worth investigating effects such as two-photon 
decay or fine and hyperfine structure when there are larger sources of 
uncertainty in predicting the 21~cm signal during the early stages of 
reionization, in particular whether or not H$_2$ cooling is active in 
low-mass haloes, the star formation efficiency, the initial mass function, 
and the X-ray luminosities of early galaxies.  Of course, answering these 
questions is a major motivation for 21~cm observations.  This paper takes 
the perspective that one can only address these questions if the 
theoretically tractable parts of the problem, such as the \wuf\ coupling 
strength, have been solved.  Otherwise, degeneracies exist in the data 
that cannot be broken, e.g. one could change the emitted UV spectra of the 
stars and also change how the \lya\ production probability $P_{np}$ 
depends on quantum level $n$.  Also, one cannot establish that an effect 
such as the fine structure of \lya\ is negligible until it has been 
calculated.

The results of this paper mostly affect the 21~cm signal during a narrow 
redshift range near the beginning of reionization.  This is because at 
earlier times there were no UV photons, so the \wuf\ effect is 
unimportant, and at later times there were so many UV photons that the 
\wuf\ effect is the only important mechanism determining the spin 
temperature, so that $T_s=T_k$ regardless of the details.  The transition 
region, in which UV photons compete with the CMB for control of the spin 
temperature, may have been brief but it is a gold mine of information on 
early galaxies.  For example, \citet{2004astro.ph.10129B} have suggested 
that the fluctuations in the UV radiation could be detectable, providing 
information about the clustering of the first stars.

This paper is organized as follows.  \S\ref{s:basic} explains the 
formalism used to predict the 21~cm signal and defines relevant notation.  
The main results of the paper are in \S\ref{s:wuf}, including the 
calculation of the probabilities for \lya\ emission and two-photon decay, 
the new calculation of the \lya\ line profile, and a fitting formula for 
the \wuf\ coupling efficiency.  \S\ref{s:model} illustrates how the 
changes in the physics affect the 21~cm signal in two toy models of 
reionization.  I conclude in \S\ref{s:conc}.

\section{High-redshift \hi\ 21 cm radiation}
\label{s:basic}

This section reviews the basic theory of the 21~cm radiation from the 
pre-reionization IGM.  More details can be found in the references.

The brightness temperature of the 21~cm signal is determined by the spin 
temperature $T_s$ of the \hi\ according to the relation (e.g. 
\citealt{2004ApJ...608..622Z})
\begin{equation}
T_b = \frac{3c^3\hbar A_{10}\nh
\xhi}{16\nu_{10}^2k_B(1+z)^2(\rmd v_\parallel/\rmd r_\parallel)}
\left( 1 - \frac{T_\gamma}{T_s} \right),
\end{equation}
where $\rmd v_\parallel/\rmd
r_\parallel$ is the physical velocity gradient at redshift $z$; $A_{10}$ 
is the intrinsic width of the $F=1$ hyperfine level; $\nu_{10}=1.42\,$GHz 
is the \hi\ hyperfine transition frequency; $\nh$ is the proper number 
density of hydrogen nuclei; $\xhi$ is the fraction of hydrogen that is 
neutral; $T_\gamma=2.73(1+z)\,$K is the CMB temperature; and $T_s$ is the 
\hi\ spin temperature.  In the 
linear regime, the velocity field is related to the matter field by
\citep{2004MNRAS.352..142B, 2005ApJ...624L..65B}
\begin{equation}
\frac{\rmd v_\parallel}{\rmd r_\parallel}
= \frac{H(z)}{1+z}\left( 1 + f \partial_{\chi} \nabla^{-2} 
\delta \right).
\end{equation}
Here $f = \rmd[\ln (D/a)]/\rmd\ln a$ depends on the growth factor and is 
equal to $f\approx 1$ in the matter-dominated era (a good approximation 
at the redshift of reionization), and $\chi$ is the comoving radial 
distance.  Plugging in numbers from the currently 
favoured cosmology gives
\begin{equation}
T_b \approx 28\,{\rm mK}
\left( \frac{1+z}{10}\right)^{1/2} \left( 1 - \frac{T_\gamma}{T_s} \right)
\left( 1 + f \partial_{\chi} \nabla^{-2} \delta \right)^{-1},
\end{equation}
Here $\chi$ is comoving radial distance.  In 
the second line the homogeneous-universe and peculiar velocity terms have 
been separated out from each other.

The spin temperature is determined by three effects: the radiative 
coupling to the CMB, and the \wuf\ and collisional coupling to the gas 
kinetic temperature $T_k$.  These effects compete to determine the 
fraction $y$ of hydrogen atoms in the $F=1$ excited hyperfine level.  This 
fraction is related to the spin temperature via
\begin{equation}
\frac{y}{1-y}=3\rme^{-T_\star/T_s} \;\;\;
\rightarrow \;\;\;
y = \frac{3}{3+\rme^{T_\star/T_s}},
\label{eq:get-y}
\end{equation}
where $T_\star = h\nu_{10}/k_B=68.2\,$mK.  Sometimes we will write the 
populations of the excited and ground hyperfine levels $y_1=y$ and 
$y_0=1-y$ for simplicity.  At $T_s\gg T_\star$, one may use the 
approximation
\begin{equation}
y \approx \frac{3}{4} - \frac{3T_\star}{16T_s},
\end{equation}
The evolution of $y$ can be broken into
its CMB, \wuf, and collisional terms,
\begin{equation}
\dot{y} = \dot{y}_\gamma + \dot{y}_\alpha + \dot{y}_c.
\label{eq:ydot}
\end{equation}
The radiative term is given by
\begin{equation}
\dot{y}_\gamma = -\frac{4A_{10}T_\gamma}{T_\star}\left( y - \frac{3}{4} +
\frac{3T_\star}{16T_\gamma} \right),
\label{eq:ydg}
\end{equation}
where $T_\gamma$ is the photon temperature.  The factor of 
$4T_\gamma/T_\star$ in front accounts for the acceleration of the 
radiative transition via stimulated emission and absorption (which 
contributes a factor of 3 since the $F=0$ state can be excited to any of 
the three $F=1$ states), 
which dominate over spontaneous emission for $T_\gamma\gg T_\star$.  The 
collisional term is
\begin{equation}
\dot y_c = -\frac{4A_{10}T_\gamma}{T_\star} x_c\left( y - \frac{3}{4} +
\frac{3T_\star}{16T_k} \right),
\end{equation}
where
\begin{equation}
x_c = \frac{\kappa_{10}\nh T_\star}{A_{10}T_\gamma}
\end{equation}
and $\kappa_{10}$ is the collisional rate coefficient 
\citep{2005ApJ...622.1356Z}.\footnote{This is roughly given by 
$\kappa_{10}\approx 4\tilde\kappa_{10}/3$, where $\tilde\kappa_{10}$ is 
the coefficient tabulated by \citet{1969ApJ...158..423A}.}
The \wuf\ rate is
\begin{equation}
\dot y_\alpha = -\frac{4A_{10}T_\gamma}{T_\star} x_\alpha
\left( y - \frac{3}{4} + \frac{3T_\star}{16T_k} \right).
\label{eq:xalpha}
\end{equation}
This is given by
\begin{equation}
x_\alpha = \frac{8\pi\lambda_\mlya^2\gamma T_\star}
{9 A_{10} T_\gamma} S_\alpha J_\alpha,
\label{eq:salpha}
\end{equation}
where $\gamma = 50\,$MHz is the half-width at half maximum (HWHM) of the 
\lya\ resonance.  Here $J_\alpha$ is the flux of \lya\ photons (in 
cm$^{-2}$~s$^{-1}$~Hz$^{-1}$~sr$^{-1}$), and $S_\alpha$ is a factor of 
order unity that accounts for spectral distortions.
\citet{2004ApJ...602....1C} provide values for $S_\alpha$ that 
are typically of order unity.  In this paper the values of 
$S_\alpha$ are revised downward slightly after accounting for several new 
processes that affect the colour temperature and spectral profile of the 
\lya\ feature.

The final spin temperature is the steady-state solution to 
Eq.~(\ref{eq:ydot}),
\begin{equation}
1 - \frac{T_\gamma}{T_s} = \frac{x_\alpha+x_c}{1+x_\alpha+x_c}
\left( 1 - \frac{T_\gamma}{T_k} \right).
\end{equation}

\section{Wouthuysen-Field coupling efficiency}
\label{s:wuf}

\lya\ photons are produced in neutral regions of the universe in one of 
two ways: either photons can be cosmologically redshifted into the \lya\ 
resonance, or they can be emitted as part of the radiative cascade to the 
\hi\ ground state following capture of a higher-order Lyman series photon.  
Once produced, \lya\ photons couple the spin temperature of \hi\ to the 
gas kinetic temperature via the \wuf\ mechanism until they are redshifted 
out of the resonance.  This section computes the \wuf\ coupling rate as a 
function of the radiation field entering each of the Lyman lines. 
\S\ref{s:prod} computes the probability that a photon entering a 
Lyman-series resonance cause a radiative cascade in the excited \hi\ atom 
that terminates with a two-photon decay from the $2s_{1/2}$ level and 
produces no \lya.  Decay of an \hi\ atom from $2s$ involves a competition 
between the two-photon process and collisions that transfer the atom to 
$2p$; only the latter yields \lya\ photons \citep{1951ApJ...114..407S, 
1955MNRAS.115..279S}.  The usual assumption is that the \lya-producing 
channels dominate, however in the IGM the opposite is true: collisions are 
negligible (see Appendix~\ref{s:2photon}).  \S\ref{s:scatrate} 
investigates the effect of fine and hyperfine structure and frequency 
changes during spin-flip events using the Fokker-Planck equation.  
Fitting formulae for these results are presented in \S\ref{ss:practical}.  
\S\ref{s:monte} tests the assumptions of the Fokker-Planck equation by 
comparing its predictions to Monte Carlo simulations that are 
computationally intensive but do not make any approximations to the 
frequency redistribution matrix.  There it is shown that the Fokker-Planck 
equation reproduces the \wuf\ coupling strength predicted by the 
simulations to within $\le 3$ per cent at $T_k\ge 2\,$K.

\subsection{\lya\ production efficiency}
\label{s:prod}

\hi\ in the IGM is normally found in its ground configuration, $1s$.  If a 
photon is emitted into the IGM at energies between the \lya\ resonance at 
10.2~eV and the Lyman edge at 13.6~eV, it redshifts cosmologically until 
it reaches one of the Lyman-series resonances.  Because the Lyman lines in 
the neutral IGM are optically thick, the photon will be absorbed and one 
\hi\ atom is boosted into the $np$ configuration ($n\ge 2$). The excited 
state is unstable and decays through a radiative cascade.  Ultimately the 
cascade ends in one of three possibilities: (a) a \lya\ photon is emitted 
from the $2p$ configuration, leaving the \hi\ atom in the ground 
configuration; (b) the \hi\ atom reaches the metastable $2s$ 
configuration; or (c) the \hi\ atom decays directly from $n'p$ (with 
$2<n'<n$) to $1s$, emitting a higher-order (Ly$\beta$, Ly$\gamma$, etc.) 
photon.  In case (c), the emitted photon immediately re-excites an \hi\ 
atom to the $n'p$ configuration; the process of absorption and re-emission 
ultimately terminates in either (a) or (b).  In case (a), the original 
photon is downgraded to \lya.  Appendix~\ref{s:2photon} shows that in 
case (b) the atom in the $2s$ configuration decays almost always via 
two-photon emission.  The latter process, of course, produces no \lya.  
Thus the \lya\ photon production rate depends on the branching fractions 
for cases (a) and (b), which are evaluated next.

[There is so much \hi\ in the early universe that some of the electric 
quadrupole lines $1s_{1/2}\rightarrow nd_{3/2,5/2}$ are optically thick.  
Since some of these lines have slightly higher energy than the electric 
dipole lines $1s_{1/2}\rightarrow np_{1/2,3/2}$ due to fine structure, one 
might worry that a photon will redshift into the quadrupole resonance 
first and excite a hydrogen atom to the $nd$ rather than $np$ 
configuration.  However a simple calculation shows that for $n\ge 3$, the 
fine structure splitting $29n^{-3}\,$GHz between $np_{3/2}$ and $nd_{5/2}$ 
levels is less than the $1.0(1-n^{-2})(T/{\rm K})^{1/2}\,$GHz Doppler 
width of the Lyman line for the temperatures $T\ge 2\,$K expected in the 
IGM.  The splitting between $np_{3/2}$ and $nd_{3/2}$ is even less, as it 
is due to Lamb shifts and hyperfine splitting.  Thus for the purposes of 
photon absorption, the $np_{3/2}$ and $nd_{3/2,5/2}$ levels are degenerate 
and absorption occurs in the stronger electric dipole line.]

\begin{figure}
\includegraphics[angle=-90,width=3.2in]{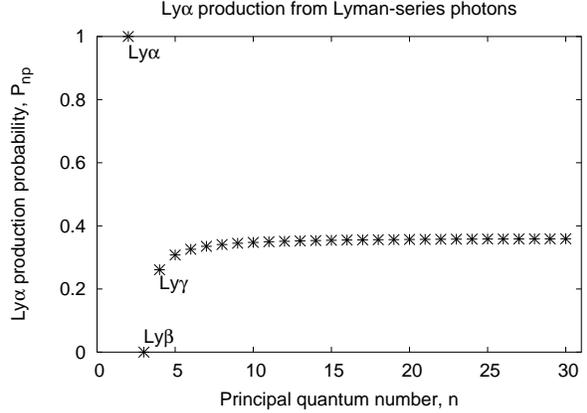}
\caption{\label{f:lyap}The probabilities $P_{np}$ of producing a \lya\ 
photon following excitation of \hi\ to the $np$ configuration.  For 
example, if a photon redshifts into the Ly$\gamma$ resonance 
($1s\rightarrow 4p$), there is a probability $P_{4p}=0.26$ that the photon 
is degraded to \lya\ and a probability $1-P_{4p}=0.74$ that it is lost to 
two-photon emission and never contributes to the \wuf\ coupling.}
\end{figure}

Let us define $P_{nl}$ to be the probabilty for an \hi\ atom in the $nl$ 
configuration to decay ultimately via \lya\ emission.  The $2s\rightarrow 
1s$ two-photon emission probability is then $1-P_{nl}$.  The probabilities 
can be determined iteratively via the usual equation:
\begin{equation}
P_{nl} = \frac{
   \sum_{n'=2}^{n-1} \sum_{l'=0}^{n'-1} A_{nl\rightarrow n'l'} P_{n'l'}
   }{
   \sum_{n'=2}^{n-1} \sum_{l'=0}^{n'-1} A_{nl\rightarrow n'l'}
   },
\label{eq:pnl}
\end{equation}
where the $A_{nl\rightarrow n'l'}$ are the decay rate coefficients (in. 
e.g. s$^{-1}$) to the specified states.  The $np\rightarrow 1s$ rate is 
removed from the sum since it results in a Lyman-series photon that 
immediately re-excites a hydrogen atom to $np$, and decays from non-$p$ 
states to $1s$ are forbidden.  Since an \hi\ atom in the $2s$ 
configuration always undergoes two-photon emission, whereas an atom in 
$2p$ undergoes \lya\ emission, Eq.~(\ref{eq:pnl}) can be initialized with 
$P_{2s}=0$ and $P_{2p}=1$.  The resulting probabilities for producing 
\lya\ photons are shown in Fig.~\ref{f:lyap} and Table~\ref{t:lyap}.  Note 
in particular that 
all photons that redshift into \lya\ end up in the \lya\ resonance 
($P_{2p}=1$), whereas none of the photons that redshift into Ly$\beta$ do 
($P_{3p}=0$) because the $3p$ configuration always decays to $1s$ or $2s$ 
on account of electric dipole selection rules.  Photons entering 
higher-order Lyman resonances can go either way ($0<P_{np}<1$).

\begin{table}
\caption{\label{t:lyap}The probabilities $P_{np}$ of producing a \lya\
photon following excitation of \hi\ to the $np$ configuration.}
\begin{tabular}{rcrcrc}
\hline\hline
$n$ & $P_{np}$ & $n$ & $P_{np}$ & $n$ & $P_{np}$ \\
\hline
 & &
11 & 0.3496 &
21 & 0.3572 \\
 2 & 1.0000 &
12 & 0.3512 &
22 & 0.3575 \\
 3 & 0.0000 & 
13 & 0.3524 &
23 & 0.3578 \\
 4 & 0.2609 &
14 & 0.3535 &
24 & 0.3580 \\
 5 & 0.3078 &
15 & 0.3543 &
25 & 0.3582 \\
 6 & 0.3259 &
16 & 0.3550 &
26 & 0.3584 \\
 7 & 0.3353 &
17 & 0.3556 &
27 & 0.3586 \\
 8 & 0.3410 &
18 & 0.3561 &
28 & 0.3587 \\
 9 & 0.3448 &
19 & 0.3565 &
29 & 0.3589 \\
10 & 0.3476 &
20 & 0.3569 &
30 & 0.3590 \\
\hline\hline
\end{tabular}
\end{table}

\subsection{Scattering rate}
\label{s:scatrate}

The efficiency of \wuf\ coupling is determined by the \lya\ spin-flip rate 
$x_\alpha$ and the degree to which the photon spectrum in the vicinity of 
\lya\ has relaxed to the gas kinetic temperature $T_k$.  It is generally 
believed that relaxation of the colour temperature to $T_k$ is complete if 
the optical depth through the \lya\ resonance (i.e. the Gunn-Peterson 
depth $\tau_{GP}$) is high enough; \citet{1985ApJ...290..578D} showed that 
if \lya\ can be treated as a single line, this relaxation occurs for 
$\tau_{GP}\ge 10^5$, which holds at all redshifts prior to reionization.  
The usual computation also assumes that each \lya\ scattering by an \hi\ 
atom in the $1s_{1/2}(F=1)$ level has a $\frac{4}{27}$ probability of 
transferring the atom to $1s_{1/2}(F=0)$, as computed by \citet{field1958, 
1959ApJ...129..536F}.

The $\frac{4}{27}$ probability was derived assuming that the $J(\nu)$ is 
constant across the \lya\ multiplet.  While this is appropriate in the 
context of ISM studies where $k_BT_k/h$ is much greater than the width of 
the \lya\ spectral feature \citep{1959ApJ...129..536F}, the 
pre-reionization IGM may have been cold, with the minimum temperature 
determined by the onset of X-ray heating \citep{2004ApJ...602....1C}.  In 
this case, the use of frequency-averaged cross sections, as in 
\citet{field1958}, is no longer valid and one must treat the line profile 
in detail.  This is done in Appendix~\ref{s:lya}, where the \lya\ line 
profile is broken into the parts $\phi_{F_iF_f}(\Delta\nu)$ that give the 
rate of scattering from initial total spin $F_i\in\{0,1\}$ to final 
$F_f\in\{0,1\}$.  Also, the \wuf\ coupling implies some transfer of energy 
between the \lya\ photons and the hydrogen spins, hence the colour 
temperature relaxes not to $T_k$ but to some value intermediate between 
$T_k$ and $T_s$.  This effect reduces $x_\alpha$ because the \wuf\ energy 
transfer rate contains the temperature difference between $T_c$ and $T_s$, 
instead of between $T_k$ and $T_s$.

This section introduces this new physics to obtain the \lya\ spectral 
distortion and to compute $x_\alpha$.  It is based on the treatment of the 
photon spectrum using the Fokker-Planck equation, which assumes that the 
change in frequency $\delta\nu$ in a single scattering event is small in 
comparison to the frequency scale over which the photon intensity or the 
scattering coefficients change.  These assumptions are not strictly valid, 
and for this reason \S\ref{s:monte} will be devoted to testing them.

A distinction is made between ``continuum'' photons that cosmologically 
redshift into \lya, and ``injected'' photons that are produced as part of 
a radiative cascade.  We find that in terms of the \wuf\ coupling, there 
is very little difference between these, in accordance with the results of 
\citet{2004ApJ...602....1C}.

The kinetic temperature range considered will be $T_k\ge 2\,$K, which 
occurs if the universe cools adiabatically until $z=9$.  In practice, 
lower temperatures were probably not reached: the universe may have been 
partially or fully reionized by $z=9$, and even inefficient heating 
sources such as \lya\ heating could have kept the universe warmer than 
$2\,$K throughout reionization (e.g. \citealt{2004ApJ...602....1C}).

\subsubsection{The \lya\ spectral distortion}
\label{s:distort}

The steady-state Fokker-Planck equation used by 
\citet{2004ApJ...602....1C} is easily 
modified to include the full (non-Voigt) line profile.  In the vicinity of 
the \lya\ 
resonance, the equation can be written as
\begin{equation}
\frac{\partial}{\partial\nu}\left( -{\cal A}J + D\frac{\partial 
J}{\partial\nu} \right) + C\psi(\nu) = \dot{J}(\nu) = 0,
\label{eq:fp}
\end{equation}
where ${\cal A}$ is the frequency drift (in Hz~s$^{-1}$), $D$ is the 
frequency diffusivity (in Hz$^2\,$s$^{-1}$), $C$ is the photon source 
term, and $\psi$ is the frequency distribution with which photons are 
injected.  The drift and diffusivity can be decomposed as
\begin{equation}
{\cal A} = {\cal A}_H + {\cal A}_k + {\cal A}_s
\label{eq:a}
\end{equation}
and
\begin{equation}
D = D_k + D_s + D_{int}.
\label{eq:d}
\end{equation}
This includes terms due to Hubble expansion (subscript $_H$), 
kinetic/Doppler coupling ($_k$), and spin coupling ($_s$).  The diffusion 
term contains an ``interference'' contribution if the diffusion due to 
kinetic coupling is correlated with that due to spin coupling; it is shown 
later in this section that $D_{int}$ can be neglected.  (Hubble expansion 
causes a drift in 
the frequency, but no diffusion.)  The Hubble expansion term is trivial, 
${\cal A}_H = -H\nu_\mlya$.

The kinetic and spin diffusion terms can be worked out from the usual 
Fokker-Planck rules, which state that for any process $X$,
\begin{equation}
{\cal A}_X = \Gamma_{\rm scat}\langle \delta\nu_X \rangle
\end{equation}
and
\begin{equation}
D_X = \frac{1}{2}\Gamma_{\rm scat}\langle \delta\nu^2_X \rangle,
\end{equation}
where $\Gamma_{\rm scat}$ is the scattering rate (in s$^{-1}$) and 
$\langle \delta\nu_X \rangle$ and $\langle \delta\nu^2_X \rangle$ are the 
mean change in frequency and mean square change in frequency during a 
scattering.  The (spin-averaged) scattering rate is
\begin{equation}
\Gamma_{\rm scat} = \frac{3}{2}\lambda^2_\mlya \gamma \nh\xhi c
  \bar\phi(\nu),
\end{equation}
where
\begin{equation}
\bar\phi(\nu) = \frac{1}{4}(\phi_{00}+\phi_{01}) + \frac{3}{4}
(\phi_{10}+\phi_{11})
\end{equation}
is the spin-averaged cross-section appropriate for $T_s\gg T_\star$;
c.f. Eq.~(\ref{eq:sigphi}).  As usual with Fokker-Planck equations, the 
drift and diffusion terms obey an Einstein relation
\begin{equation}
{\cal A}_X = -\frac{h}{k_BT_X}D_X,
\label{eq:einstein}
\end{equation}
where $T_X$ is the temperature of the reservoir with which the photon 
exchanges energy during process $X$.  Here $X$ is either $k$ (Doppler 
shift, for which $T_k$ appears in Eq.~\ref{eq:einstein}) or $s$ (spin 
coupling with $T_s$).  Physically, the kinetic drift term ${\cal A}_k$ 
corresponds to the loss of photon energy due to atomic recoil.  The 
spin drift term ${\cal A}_s$ corresponds to the loss of photon energy due 
to having more atoms in the $F=0$ than $F=1$ level, so that if the photon 
spectrum were flat ($\rmd J/\rmd\nu=0$) the photons would on average 
lose more energy in spin-flip excitations than they gain in 
de-excitations.

The kinetic diffusion has been worked out by
\citet{1994ApJ...427..603R}\footnote{\citet{1994ApJ...427..603R} work in 
terms of the variable $x$, which is related to the detuning by 
$\Delta\nu = \sqrt{2}\sigma_\nu x$.  In this paper, including 
Eq.~(\ref{eq:dk}), I have converted to $\Delta\nu$.}, with the result 
that $\langle\delta\nu^2_k\rangle = 2\sigma_\nu^2$ and hence
\begin{equation}
D_k = \frac{3}{2}\lambda_\mlya^2\gamma\nh\xhi c\sigma_\nu^2 \bar\phi(\nu),
\label{eq:dk}
\end{equation}
where
$\sigma_\nu$ is the $1\sigma$ Doppler width.  [In the Fokker-Planck 
approximation, and for an isotropic situation, the angular dependence of 
the cross section enters into Eq.~(\ref{eq:dk}) only through the 
combination $\langle 1-{\bmath n}\cdot{\bmath n}'\rangle$ 
where ${\bmath n}$ and ${\bmath n}'$ are the incoming and outgoing photon 
directions; see Eqs.~(A15) and (A16) of \citet{1994ApJ...427..603R}.  So 
long as only the electric dipole transitions are involved in scattering, 
the probability for scattering into direction ${\bmath n}'$ is the same 
as for $-{\bmath n}'$, and the angular dependence requires no 
modification to Eq.~(\ref{eq:dk}).]  Equation (\ref{eq:einstein}) then 
gives ${\cal A}_k$.

\citet{2004ApJ...602....1C} included in their Fokker-Planck equation only 
the Hubble drift, kinetic drift, and kinetic diffusivity (in their Eq.~13, 
the Hubble drift is the $\gamma_S$ term and the kinetic drift is the 
$\eta$ term).  However the hyperfine splitting of the ground 
state allows a photon to change its frequency during scattering by 
$\pm\nu_{10}$, even in the centre-of-mass frame.  This results in spin 
contributions to the drift and diffusivity.  Spin diffusivity results only 
from those \lya\ scattering events that change the total spin state of the 
atom; in the limit $T_s\gg T_\star$,
\begin{equation}
D_s = \frac{3}{4}\lambda_\mlya^2\gamma\nh\xhi c\nu_{10}^2
\left( \frac{1}{4}\phi_{01} + \frac{3}{4}\phi_{10} \right).
\label{eq:ds}
\end{equation}
Equation (\ref{eq:einstein}) can then be used to obtain ${\cal A}_s$.

Finally there is the interference diffusivity $D_{int}$ in 
Eq.~(\ref{eq:d}).  This term comes from the fact that one cannot exactly 
separate $\langle\delta\nu^2\rangle$ into kinetic and spin parts, 
$\langle\delta\nu^2_k\rangle+\langle\delta\nu^2_s\rangle$, and is equal to 
\begin{equation}
D_{int} = \Gamma_{\rm scat} \langle \delta\nu_k \delta\nu_s \rangle.
\end{equation}
In our particular case, the deviation of $\delta\nu_k$ from its mean value 
$\langle\delta\nu_k\rangle$ is proportional to ${\bmath n}\cdot{\bmath 
n}'$.  However as argued above, the probabilities of scattering the 
photon in directions ${\bmath n}'$ and $-{\bmath n}'$ are equal; the same 
argument holds for the conditional probabilities for fixed final spin $F_f$.
Therefore $\delta\nu_k$ is uncorrelated with $\delta\nu_s$, and $D_{int} = 
\Gamma_{\rm scat} \langle \delta\nu_k \rangle\langle\delta\nu_s \rangle$.  
Combining with Eqs.~(\ref{eq:a}), (\ref{eq:d}), and (\ref{eq:einstein}) 
shows that
\begin{equation}
\frac{D_{int}}{\sqrt{D_kD_s}} =
\frac{h\langle\delta\nu_k^2\rangle^{1/2}}{k_BT_k}
\frac{h\langle\delta\nu_s^2\rangle^{1/2}}{k_BT_s}.
\label{eq:di}
\end{equation}
It is readily verified that $h\langle\delta\nu_X^2\rangle^{1/2}\ll k_BT_X$ 
for $X=k,s$ so long as
$T_k \gg (h\nu_\mlya)^2/k_Bm_pc^2 = 1.3\,$mK
and $T_s\gg T_\star$, respectively.  Both of these conditions are easily 
satisfied in the IGM, and so $D_{int}\ll \sqrt{D_kD_s}$.  Since 
it is strictly true that $\sqrt{D_kD_s}\le (D_k+D_s)/2$, 
$D_{int}$ can be dropped.\footnote{It is a good thing that $D_{int}$ can 
be neglected, 
since if we had $h\langle\delta\nu_X^2\rangle^{1/2}\ge k_BT_X$ then the 
typical change in frequency in a single scattering would be comparable to 
$k_BT_X/h$, i.e. to the width of the spectral features.  In this case the 
assumptions of the Fokker-Planck equation would not be valid.}

Equation (\ref{eq:fp}) can be solved by the method of
\citet{2004ApJ...602....1C}, which consists of first reducing it to first 
order,
\begin{eqnarray}
-{\cal A}(\nu)J(\nu) + D(\nu)\frac{\partial J(\nu)}{\partial\nu} &=&
-{\cal A}(-\infty)J(-\infty)
\nonumber \\ && - C\int_{-\infty}^\nu \psi(\nu')\rmd\nu',
\label{eq:j1}
\end{eqnarray}
and then applying an ordinary differential equation (ODE) solver starting 
from $\nu=-\infty$ and working upward in frequency.\footnote{The initial 
condition is technically given by $J(-\infty)=J_\alpha$.  In 
practice, any errors in the initial condition of Eq.~(\ref{eq:j1}) are 
damped as $\propto \exp \int [{\cal A}/D]\rmd\nu$.  Since ${\cal 
A}\le{\cal A}_H<0$ and the diffusivity $D\rightarrow 0$ far from 
resonance, all solutions of Eq.~(\ref{eq:j1}) rapidly converge to the
physical solution if the integration is initiated at sufficiently 
negative $\Delta\nu$.}  Here $J(-\infty)=J_\alpha$ is the total flux at 
\lya\ (including both continuum and injected photons) and $C$ is 
determined only by the injected photons.  One can determine $C$ as 
follows: substituting $\nu=+\infty$ into Eq.~(\ref{eq:j1}), and recalling 
that $\psi$ integrates to unity, one finds
\begin{equation}
-{\cal A}(+\infty)J(+\infty) = -{\cal A}(-\infty)J(-\infty) - C,
\end{equation}
implying
\begin{equation}
C = -{\cal A}(-\infty)[J(+\infty)-J(-\infty)]
= H\nu_\mlya J_\alpha({\rm inj}).
\end{equation}

\subsubsection{Effect on spin temperature}

Once a solution to Eq.~(\ref{eq:fp}) is obtained, one can go back and 
estimate the \wuf\ effect on the spin temperature.  The rate 
per atom
$\Gamma_{\alpha 10}$ for converting $F=1$ hydrogen atoms to $F=0$ is
\begin{eqnarray}
\Gamma_{10} &=& 4\pi \int J(\nu)\sigma(1\rightarrow 0;\nu)\;\rmd\nu
\nonumber \\
&=& 6\pi\lambda_\mlya^2\gamma J_\alpha \int \frac{J(\nu)}{J_\alpha}
\phi_{10}(\nu)\,\rmd\nu,
\label{eq:gamma10}
\end{eqnarray}
and a similar rate holds for $F=0\rightarrow 1$ conversions.  If  
$y$ is the fraction of hydrogen atoms in the excited hyperfine level 
$F=1$, then the \wuf\ contribution to $\dot{y}$ is
\begin{equation}
\dot{y}_\alpha = (1-y)\Gamma_{01} - y\Gamma_{10}
= (\Gamma_{01}+\Gamma_{10})\left( y - y_{\alpha,ss} \right),
\label{eq:yalpha}
\end{equation}
where
\begin{equation}
y_{\alpha,ss} = \frac{\Gamma_{01}}{\Gamma_{01}+\Gamma_{10}}
\label{eq:yass}
\end{equation}
is the steady-state occupation fraction of the excited level if the
\wuf\ effect were the only effect operating and if the \lya\ spectral 
shape were fixed.  For the special case where 
the photon spectrum is thermal across the \lya\ line with colour 
temperature $T_c$, $J(\nu)\propto 
\exp(-h\nu/k_BT_c)$, one would have $y_{\alpha,ss}=3/4-3T_\star/16T_c$.  
In reality the spectrum in the vicinity of the \lya\ resonance is 
non-thermal, and the effective colour temperature $-(h/k_B)\rmd\ln 
J/\rmd\nu$ is between $T_k$ and $T_s$.  However $y_{\alpha,ss}$ as defined 
by Eq.~(\ref{eq:yass}) still exists.  One can therefore define an 
effective colour temperature $T_c^{eff}$ by
\begin{equation}
\rme^{-T_\star/T_c^{eff}}\equiv \frac{y_{\alpha,ss}}{3(1-y_{\alpha,ss})} 
\;\;\;\rightarrow\;\;\;
y_{\alpha,ss} \approx \frac{3}{4} - \frac{3T_\star}{16T_c^{eff}}.
\label{eq:tceff}
\end{equation}
Then comparison of Eq.~(\ref{eq:yalpha}) with Eq.~(\ref{eq:xalpha}) yields
\begin{equation}
x_\alpha = \frac{(\Gamma_{01}+\Gamma_{10})T_\star}{4A_{10}T_\gamma}
\frac{T_c^{eff\,-1}-T_s^{-1}}{T_k^{-1}-T_s^{-1}}.
\end{equation}
This is the equation used to determine $x_\alpha$.  Note that 
$x_\alpha$ depends on all three temperatures $T_k$, $T_s$, and $T_\gamma$, 
both explicitly and through the dependence on $\Gamma_{01}$, 
$\Gamma_{10}$, and $T_c^{eff}$.  The explicit dependence on the radiation 
temperature can be eliminated by using Eq.~(\ref{eq:salpha}) to write
\begin{equation}
S_\alpha = \frac{9(\Gamma_{01}+\Gamma_{10})}
{32\pi\lambda^2_\mlya\gamma J_\alpha}
\frac{T_c^{eff\,-1}-T_s^{-1}}{T_k^{-1}-T_s^{-1}}.
\label{eq:sacompute}
\end{equation}
The value of $S_\alpha$ thus depends only on $T_k$, $T_s$, the injection 
profile $\psi(\nu)$, $H$, and $\nh\xhi$.  It does not depend on $J_\alpha$ 
because of the linearity of Eq.~(\ref{eq:fp}).  Furthermore, if one 
multiplies both $H$ and $\nh\xhi$ by some scaling factor $\beta$ while 
holding $J_\alpha$ fixed, then ${\cal A}$, $D$, and $C$ are all 
multiplied by $\beta$, hence the solution to Eq.~(\ref{eq:fp}) and the 
value 
of $S_\alpha$ are unchanged.  Therefore $S_\alpha$ can really be written 
purely as a function of $T_k$, $T_s$, $\psi(\nu)$, and the Gunn-Peterson 
depth \citep{1965ApJ...142.1633G}
\begin{equation}
\tau_{GP} = \frac{3\nh\xhi\lambda^3_\mlya\gamma}{2H},
\label{eq:tgp}
\end{equation}
which differs from the ratio $\nh\xhi/H$ only by fundamental constants.

\begin{figure}
\includegraphics[angle=-90,width=3.2in]{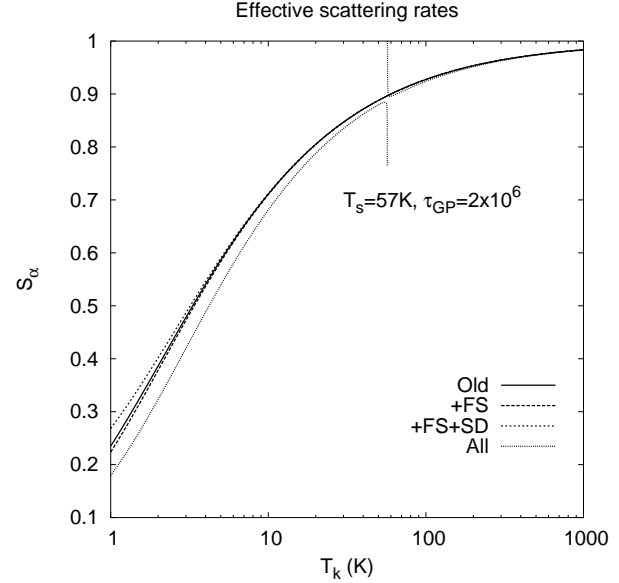}
\caption{\label{f:salpha}The \wuf\ effective coupling $S_\alpha$.  The 
solid curve shows the old calculation, which treats the \lya\ resonant 
cross section as a single Voigt profile and assumes $T_c^{eff}=T_k$.  The 
long-dashed curve (``+FS'') shows what happens when one includes the fine 
and hyperfine structure of the \lya\ line profile using 
Eq.~(\ref{eq:phiff}).  The short-dashed curve (``+FS+SD'') also includes 
the spin diffusivity, i.e. the change of frequency of a photon when it 
scatters a hydrogen atom and flips the spin state.  Finally, the dotted 
curve (``all'') represents the full calculation and includes the correct 
colour temperature $T_c^{eff}$ instead of assuming that it is completely 
relaxed to the kinetic temperature $T_k$.  Note that all of the effects 
are most important at low $T_k$.  (See the text for a discussion of the 
singularity in the ``all'' curve at $T_k=57\,$K.)}
\end{figure}

As an example, Fig.~\ref{f:salpha} shows the values of $S_\alpha$ for the 
particular case of $T_s=57\,$K and $\tau_{GP}=2\times 10^6$, which are 
reasonable for redshifts $z\approx 20$ in late-reionization scenarios 
where the \wuf\ coupling is still weak (i.e. $x_\alpha\ll 1$).  The figure 
shows both the ``old'' calculation, which neglected fine structure and 
spin diffusivity, and assumed $T_c^{eff}=T_k$, and the ``new'' calculation 
which includes fine structure and spin diffusivity and accounts for 
incomplete relaxation of the photon spectrum ($T_c^{eff}\neq T_k$).
The feature in the ``all'' curve at 
$T_k=57\,$K represents the fact that the denominator in 
Eq.~(\ref{eq:sacompute}) has a singularity when $T_k=T_s$, since even in 
this case, the Hubble expansion term in the Fokker-Planck equation implies 
that $T_c^{eff}$ is not exactly equal to $T_k$.  Because this feature 
corresponds only to a small change in $T_c^{eff}$ its has no important 
physical consequences, rather it just an annoying feature of the variable 
$S_\alpha$.  In \S\ref{ss:practical} I introduce a modified variable 
$\tilde S_\alpha$ that avoids any singular behaviour.

\subsection{Practical calculation}
\label{ss:practical}

The scattering function $S_\alpha$ is convenient conceptually, however in 
actual computation the presence of $T_k^{-1}-T_s^{-1}$ in the denominator 
is problematic.  This problem is solved by splitting $S_\alpha$ into two 
parts,
\begin{equation}
S_\alpha = \tilde S_\alpha 
\frac{T_c^{eff\,-1}-T_s^{-1}}{T_k^{-1}-T_s^{-1}},
\end{equation}
where
\begin{equation}
\tilde S_\alpha =
\frac{9(\Gamma_{01}+\Gamma_{10})}{32\pi\lambda^2_\mlya\gamma J_\alpha}
= \frac{27}{16}\int \frac{J(\nu)}{J_\alpha}
  [\phi_{10}(\nu)+\phi_{10}(\nu)]\,\rmd\nu.
\label{eq:tsa}
\end{equation}
Here $\tilde S_\alpha$ and $T_c^{eff}$ are functions of $\tau_{GP}$, 
$T_s$, and $T_k$, and the second equality uses 
Eq.~(\ref{eq:gamma10}).  One can define a modified \lya\ coupling 
parameter 
$\tilde x_\alpha$ by
\begin{equation}
\tilde x_\alpha = \frac{8\pi\lambda^2_\mlya\gamma T_\star}
{9A_{10}T_\gamma}\tilde S_\alpha J_\alpha.
\label{eq:txa}
\end{equation}
The overall spin temperature is then given by
\begin{equation}
T_s^{-1} = \frac{T_\gamma^{-1} + \tilde x_\alpha T_c^{eff\,-1} + x_c 
T_k^{-1}}
{1 + \tilde x_\alpha + x_c}.
\label{eq:spintemp}
\end{equation}
Note that since $\tilde S_\alpha$ and $T_c^{eff}$ are functions of $T_s$ 
as well as $T_k$ and $\tau_{GP}$, Eq.~(\ref{eq:spintemp}) is an implicit 
equation for the spin temperature.  The dependence is however weak, so a 
simple and robust way to find $T_s$ for given $T_\gamma$, $T_k$, 
$J_\alpha$, and $\tau_{GP}$ is to iteratively compute $T_c^{eff}$ and 
$\tilde S_\alpha$ for some value of $T_s$, and then update $T_s$ using 
Eq.~(\ref{eq:spintemp}).  Initializing the 
iteration with $T_s({\rm init}) = T_\gamma$ results in convergence to 
better than 1 per cent after $<5$ iterations for reasonable values of 
$T_k$ ($T_k>1\,$K).

The functions $\tilde S_\alpha$ and $T_c^{eff}$ cannot be computed in 
closed analytic form, and can be expensive to evaluate numerically as they 
require solution of an ODE.  Therefore the simplest method to obtain them 
is to first compute values on a grid of points in
$(\tau_{GP}, T_s, T_k)$, and then build a fitting formula.
The following formula for $\tilde 
S_\alpha$ reproduces our numerical results to within 1 per cent in the 
range $T_k\ge 2\,$K, $T_s\ge 2\,$K, and $10^5\le \tau_{GP}\le 10^7$ for 
continuum photons:
\begin{eqnarray}
\tilde S_\alpha &=&
(1 - 0.0631789T_k^{-1} + 0.115995T_k^{-2}
\nonumber \\ &&
- 0.401403T_s^{-1}T_k^{-1}
+ 0.336463T_s^{-1}T_k^{-2})
\nonumber \\ &&
\times (1 + 2.98394\xi + 1.53583\xi^2 + 3.85289\xi^3)^{-1},
\label{eq:fit-s}
\end{eqnarray}
where
\begin{equation}
\xi = (10^{-7}\tau_{GP})^{1/3}T_k^{-2/3}
\end{equation}
and the temperatures $T_k$ and $T_s$ are in Kelvins.  A simpler formula 
holds for $T_c^{eff}$ over the same range,
\begin{equation}
T_c^{eff\,-1} = T_k^{-1} + 0.405535T_k^{-1}(T_s^{-1}-T_k^{-1}),
\label{eq:fit-t}
\end{equation}
where again $T_k$ is in Kelvins.  This reproduces the $T_c^{eff\,-1}$ 
values from the Fokker-Planck equation to 1 per cent.

\begin{figure}
\includegraphics[angle=-90,width=3.2in]{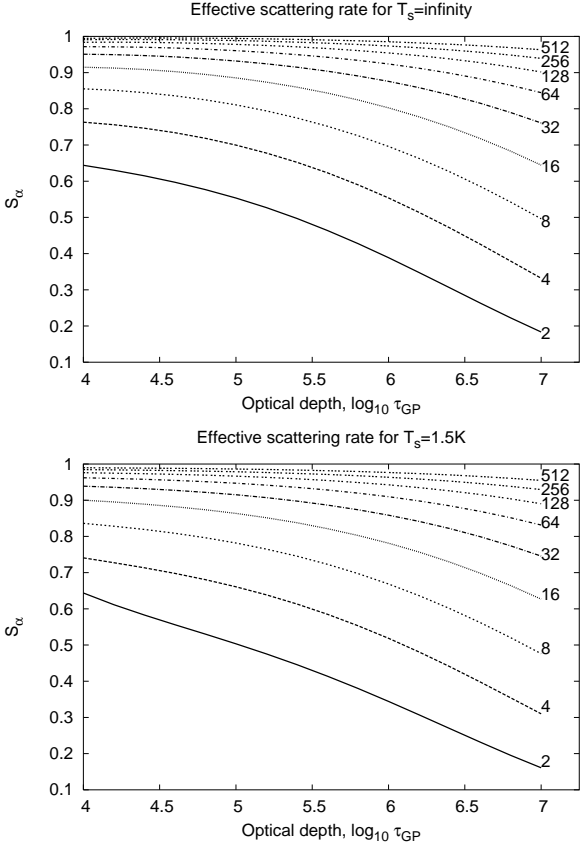}
\caption{\label{f:scat}The values of $\tilde S_\alpha$ for $T_s=\infty$ 
(top
panel) and $2\,$K (bottom panel) for continuum photons.  The curves are 
drawn for 11 values of 
$T_k$ spaced logarithmically from $1$--$10^4\,$K with intervals of 
$10^{0.4}$.  The labels shown are the values of $\log_{10}T_k$, with 
$T_k$ in Kelvins.}
\end{figure}

The numerically computed (i.e. not from the fitting formula) function
$\tilde S_\alpha$ is shown in Fig.~\ref{f:scat}.

For the injected photons, it is found that Eq.~(\ref{eq:fit-s}) reproduces 
the Fokker-Planck results for $\tilde S_\alpha$ to better than 3 per cent.  
Eq.~(\ref{eq:fit-t}) reproduces the colour temperature $T_c^{eff}$ to 
better than 4 per cent at $T_k<10^3\,$K.  At higher temperatures 
$10^3<T_k<10^4\,$K, the error increases to 12 per cent with the fitting 
formula underestimating the colour temperature.  This is because for the 
very high temperatures the photon spectrum is essentially flat, with the 
slope $T_c^{eff\,-1}$ being very close to zero.  The absolute error in 
$T_c^{eff\,-1}$ at $T_k>10^3\,$K is never greater than $3.7\times 
10^{-5}\,$K$^{-1}$, which is $<1$ per cent of $T_\gamma^{-1}$ at all 
redshifts of interest.

\subsection{Monte Carlo simulations}
\label{s:monte}

In \S\ref{s:scatrate}, we solved for the \lya\ spectral distortion 
assuming the Fokker-Planck equation to be valid.  This equation rests on 
several assumptions whose validity must be considered and tested.  The 
simplest way to test the assumptions is to use a Monte Carlo simulation, 
which is done in this section.

\citet{2004ApJ...602....1C} argued that the Fokker-Planck equation is 
valid whenever the scale in frequency over which the photon spectrum 
varies is much greater than $\sigma_\nu$.  This is true in the damping 
tails of the \lya\ resonances, but not in the Doppler cores since one must 
also drop the derivatives of the line profiles $\phi_{F_iF_f}$, as was 
done in deriving Eq.~(A15) of \citet{1994ApJ...427..603R}.  The \lya\ 
Doppler core extends out to $\ge 3.3\sigma_\nu$ at $T_k\ge 2\,$K, so 
within this region the Fokker-Planck equation does not reproduce the 
frequency redistribution matrix.  Of course, for the case considered by 
\citet{2004ApJ...602....1C} the portion of the spectrum near line centre 
is in thermal equilibrium with the atoms, with colour temperature $T_k$.  
If equilibrium applies, the correct solution is obtained regardless of the 
frequency redistribution matrix.  In this paper, however, we have 
introduced spin diffusivity for which one may have $T_s\neq T_k$, and 
thermal equilibrium does not apply.  At high kinetic temperatures this is 
irrelevant because the change in frequency $\sim\nu_{10}$ due to spin-flip 
events is negligible compared to the change $\sim\sigma_\nu$ due to the 
Doppler effect, and the photons equilibriate at colour temperature $T_k$.  
But at low kinetic temperature if $T_s\neq T_k$ no such equilibrium 
occurs, the exact form of the frequency redistribution matrix matters, and 
the validity of the Fokker-Planck equation must be verified.

The results from the Fokker-Planck equation can be checked in two basic 
ways: one could construct the integro-differential equations for $J(\nu)$ 
and solve them, or one could do a Monte Carlo simulation in which the 
distribution $J(\nu)$ is sampled rather than explicitly represented as a 
function.  In our case, the inclusion of fine/hyperfine structure and 
spin-flip (Raman) scattering makes the redistribution matrix much more 
complicated than the ``$R_{\rm II}$'' form of \citet{1985ApJ...290..578D} 
or \citet{1994ApJ...427..603R}, so the Monte Carlo method is used here.

\subsubsection{Methodology}

The basic procedure for the Monte Carlo simulation is:
\newcounter{LC1}
\begin{list}{\arabic{LC1}.~}{\usecounter{LC1}}
\item Start a photon at some starting frequency $\nu=\nu_{\rm start}$.
\item \label{i:dt}
Determine the optical depth $\delta\tau$ through which the photon 
travels before it scatters by selecting it from an exponential 
distribution: $P(\delta\tau)\,\rmd\delta\tau = 
\rme^{-\delta\tau}\,\rmd\delta\tau$.
\item Determine the photon's frequency $\nu^{(1)}$ when it scatters by 
solving the equation,
\begin{equation}
\delta\tau = \tau_{GP} \int_{\nu^{(1)}}^\nu
\sum_{F_i,F_f} y_{F_i}\phi_{F_iF_f}(\nu') \,\rmd\nu'.
\label{eq:dtmc}
\end{equation}
The Gunn-Peterson depth $\tau_{GP}$ normalizes the total optical depth.  
The Doppler-convolved line profiles $\phi_{F_iF_f}$ appear in 
Eq.~(\ref{eq:dtmc}).  If the optical depth $\delta\tau$ is not reached by 
the time the integration reaches a terminating frequency 
$\nu^{(1)}=\nu_{\rm term}$, the simulation is stopped.
\item When the photon scatters off an H atom, choose the initial and final 
spin states of the H atom.  The probability for $F_i\rightarrow F_f$ 
scattering is 
$y_{F_i}\phi_{F_iF_f}/\sum_{F'_iF'_f}y_{F'_i}\phi_{F'_iF'_f}$.
\item \label{i:u}
Once the initial and final spin states are selected, one must obtain 
the velocity ${\bmath v}$ of the atom that does the scattering.  It is 
most convenient to express this velocity in frequency units, ${\bmath 
u}=\nu_\mlya{\bmath v}/c$.  The 
component parallel to the initial direction of propagation of the photon 
we will denote $u_\parallel$.  Its probability distribution is
\begin{equation}
P(u_\parallel)\,\rmd u_\parallel =
\frac{
\rme^{-u_\parallel^2/2\sigma_\nu^2}\phi^u_{F_iF_f}(\nu^{(1)}-u_\parallel)
\,\rmd u_\parallel
}{
\int_{-\infty}^\infty
\rme^{-{u'_\parallel}^2/2\sigma_\nu^2}\phi^u_{F_iF_f}(\nu^{(1)}-u'_\parallel)
\,\rmd u'_\parallel
},
\label{eq:pu}
\end{equation}
where the $^u$ superscript denotes the un-convolved line profile.
The perpendicular component in the plane of scattering (i.e. containing 
the initial and final directions of the photon) is $u_\perp$ and has a 
Gaussian probability distribution with zero mean and variance 
$\sigma_\nu^2$.  For a Maxwellian velocity distribution, $u_\parallel$ and 
$u_\perp$ are independent.
\item \label{i:chi}
Obtain the scattering angle $\chi$, i.e. the angle between the 
incoming and outgoing photon directions.  This is obtained via
\begin{equation}
P(\chi)\,\rmd\chi = \frac{1}{2}\left[ 1 + 
5\varpi_2\left(\frac{3}{2}\cos^2\chi-\frac{1}{2}\right) \right]
\sin\chi\,\rmd\chi
\end{equation}
within the range $0\le\chi\le\pi$;
c.f. Eq.~(\ref{eq:varpi}).  The phase function $\varpi_2$ is evaluated 
using Eq.~(\ref{eq:varpi-eval}) at 
the frequency in the atom's frame, $\nu^{(1)}-u'_\parallel$, instead of 
$\nu^{(1)}$.
\item The photon's post-scattering frequency is determined by conservation 
of energy.  Specifically, the atom picks up a recoil velocity 
$\delta{\bmath v}$ with components $\delta v_\parallel = 
(1-\cos\chi)h\nu_\mlya/m_pc$ and $\delta v_\perp = 
(\sin\chi)h\nu_\mlya/m_pc$.  Its kinetic energy then changes by 
$m_p{\bmath v}\cdot\delta{\bmath v}+\frac{1}{2}m_p|\delta{\bmath v}|^2$.  
The atom also 
changes its hyperfine enrgy by $(F_f-F_i)h\nu_{10}$.  Putting these 
results together implies a post-scattering frequency
\begin{equation}
\nu^{(2)} = \nu^{(1)} - (F_f-F_i)\nu_{10} - (u_\parallel+\eta)(1-\cos\chi)
- u_\perp\sin\chi,
\end{equation}
where $\eta=h\nu_\mlya^2/m_pc^2$.
\item Replace $\nu:=\nu^{(2)}$ and return to step \#\ref{i:dt}.
\end{list}

The Monte Carlo method is straightforward in concept; the major 
non-trivial aspect is the construction of random numbers.  The 
distribution of $\delta\tau$ in step \#\ref{i:dt} and that of $u_\perp$ in 
step \#\ref{i:u} are exponential and Gaussian respectively and are 
computed using the Numerical Recipes {\tt expdev} and {\tt gasdev} 
functions \citep{1992nrca.book.....P}.  The distribution of $\chi$ in step 
\#\ref{i:chi} is also straightforward: the variable $\mu=\cos\chi$ is in 
the range $-1\le\mu\le +1$, and a simple rejection method with a constant 
comparison function (e.g. \S 7.3 of \citealt{1992nrca.book.....P}) works 
very efficiently.  The challenge is the distribution of $u_\parallel$ in 
step \#\ref{i:u} because in most cases the distribution is polymodal with 
$P(u_\parallel)$ sometimes varying by several orders of magnitude between 
the very narrow resonance peaks.  This algorithm is presented in 
Appendix~\ref{app:rng}.

The starting and terminating frequencies also require some work.  There 
are two requirements on these.  First, one does not want to miss the 
spin-flip scattering events that can occur in the \lya\ damping wings; and 
second, one does not want to artificially terminate photons that reach 
$\nu_{\rm term}$ that in reality would be scattered back to line centre.  
The first issue can be addressed by considering the number of spin-flip 
scatterings that occur in the damping wings.  Using Eq.~(\ref{eq:phiff}), 
we can find the integrated spin-flip cross section in the far damping 
wings.  For example, for $F=0\rightarrow 1$ scattering,
\begin{equation}
\int_{\nu_A+750\,\rm GHz}^\infty 
\!\!\!\!\!\!\!\!\!\!\phi_{01}^u(\nu')\,\rmd\nu'
\approx \int_{-\infty}^{\nu_A-750\,\rm GHz} 
\!\!\!\!\!\!\!\!\!\!\phi_{01}^u(\nu')\,\rmd\nu'
\approx 3\times 10^{-10};
\label{eq:intphi}
\end{equation}
the corresponding value for $1\rightarrow 0$ scattering is $1\times 
10^{-10}$.  Thus the fraction of the spin-flip events that occur more than 
750~GHz from resonance can be neglected.  The Doppler smearing does not
change this conclusion since at the temperatures of interest, 
$\sigma_\nu\ll 750\,$GHz and hence the spin-flip cross 
sections more than 750~GHz from resonance are not 
significantly affected by the Doppler effect.\footnote{The values of a 
few$\times 10^{-10}$ are several orders of magnitude less than one would 
calculate for a Lorentzian profile.  This is because the spin-flip 
process can proceed through either the $2p_{1/2}(F=1)$ or $2p_{3/2}(F=1)$ 
hyperfine excited levels.  The amplitudes through each of the excited 
levels add coherently, and they undergo destructive interference when one 
is far from resonance.}  We thus use $\nu_{\rm start}=\nu_A+750\,$GHz.

We next consider the possibility of a photon reaching $\nu_{\rm 
term}=\nu_A-1\,$THz and scattering back to line centre.  A simple way of 
evaluating how important this is is to go to the Fokker-Planck equation 
(which is valid in the damping tails) and injecting photons at the 
frequency $\nu_{\rm term}$ instead of at line centre.  Even in the worst 
case used in the Monte Carlo simulations below ($T_k=10\,$K, $T_s=\infty$, 
and $\tau_{GP}=10^6$), this gives $\tilde S_\alpha = 1.0\times 10^{-12}$, 
which implies that photons that pass through $\nu_{\rm term}$ and then 
scatter contribute this amount to the scattering rate.  Since this is 
negligible, we conclude that for the parameters simulated, $\nu_{\rm 
term}=\nu_A-1\,$THz is an acceptable terminating frequency.

Once the Monte Carlo simulation has been run, one can construct the 
quantities $\tilde S_\alpha$ and $T_c^{eff}$ as follows.  Suppose that 
during the 
course of the simulation, one observes $N_{F_iF_f}$ of the $F_i\rightarrow 
F_f$ scattering events.  The rate per unit volume (i.e. in 
cm$^{-3}\,$s$^{-1}$) at which photons are 
redshifting into the \lya\ resonance is
\begin{equation}
\dot{n}_\gamma = 4\pi H\nu_\mlya c J_\alpha,
\end{equation}
where the factor of $4\pi c$ converts the ``per unit area 
per unit time per unit solid angle'' in the definition of $J_\alpha$ into 
``per unit volume,'' and $H\nu_\mlya$ is the rate at 
which the photon's frequency is changing.  The rate of $F_i\rightarrow
F_f$ scattering events per neutral atom in the $F_i$ level is then
\begin{equation}
\Gamma_{F_iF_f} = \frac{\dot{n}_\gamma\langle N_{F_iF_f}\rangle}{\nh\xhi 
y_{F_i}}
\label{eq:gfiff}
\end{equation}
(this has units of s$^{-1}$).  Comparison to Eq.~(\ref{eq:gamma10}), and 
use of Eq.~(\ref{eq:tgp}) to express the Hubble rate and the number 
densities in terms of $\tau_{GP}$, yields the expression
\begin{equation}
\int \frac{J(\nu)}{J_\alpha}\phi_{F_iF_f}(\nu)\,\rmd\nu = 
\frac{\langle N_{F_iF_f}\rangle}{\tau_{GP} y_{F_i}}.
\label{eq:n-to-j}
\end{equation}
Equation~(\ref{eq:n-to-j}) allows us to obtain $\tilde S_\alpha$ by 
plugging the results into Eq.~(\ref{eq:tsa}).  One may also obtain the 
colour temperature by plugging the rates (Eq.~\ref{eq:gfiff}) into 
Eqs.~(\ref{eq:yass}) and (\ref{eq:tceff}); the result is
\begin{equation}
\rme^{-T_\star/T_c^{eff}} = \frac{
y_0 \langle N_{10}\rangle
}{
3 y_1 \langle N_{01}\rangle
}.
\label{eq:n-to-t}
\end{equation}

Error estimates on $\tilde S_\alpha$ and $T_c^{eff}$ may be computed by 
taking the covariance matrix of $N_{10}$ and $N_{01}$, obtained from the 
dispersion among many Monte Carlo simulations, and propagating these to 
$\tilde S_\alpha$ and $T_c^{eff}$ using the usual Jacobian rules.

\subsubsection{Results}

\begin{figure*}
\includegraphics[angle=-90,width=6.5in]{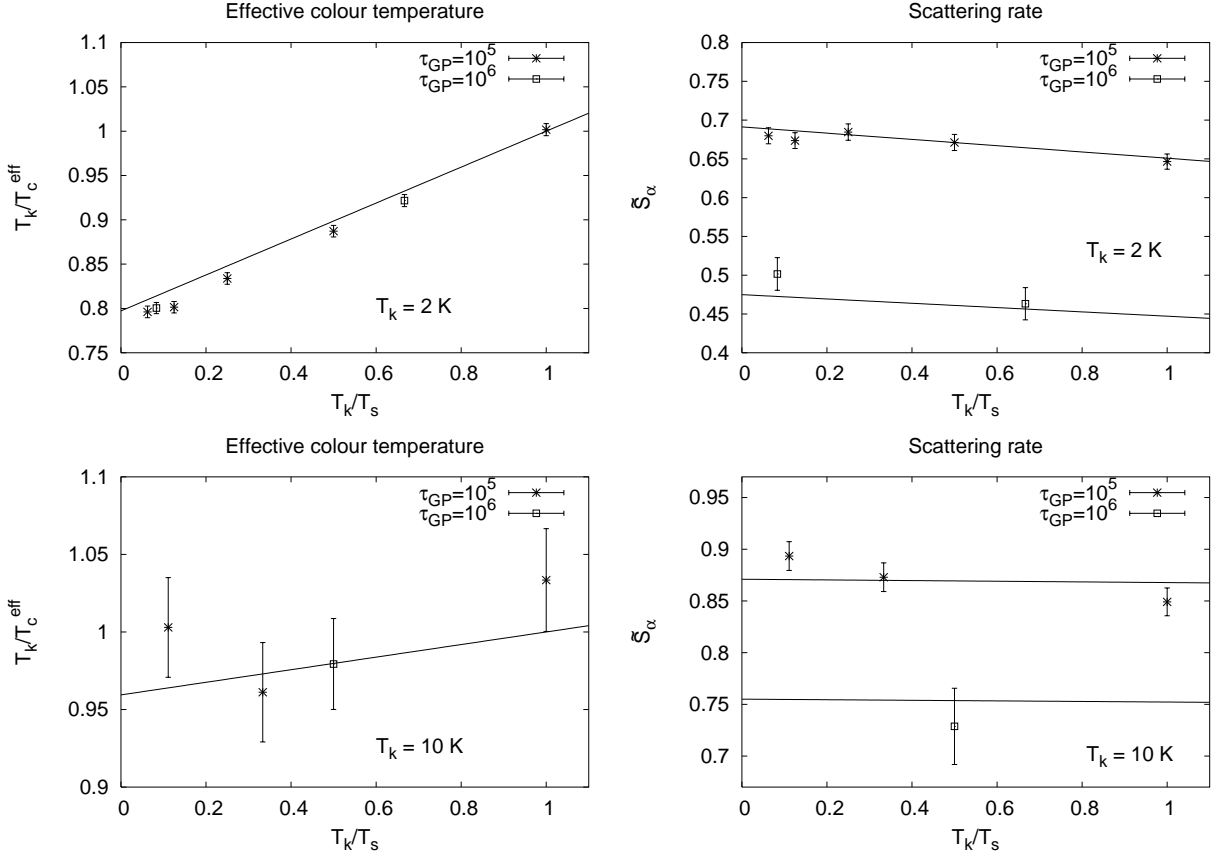}
\caption{\label{f:matrix}The colour temperatures and spin-flip scattering 
rates for several values of $T_k$ and $\tau_{GP}$.  The points with error 
bars come from Monte Carlo simulations at the specified values of 
$T_k/T_s$ and $\tau_{GP}$.  The curves in the left panel come from 
Eq.~(\ref{eq:fit-s}), and those in the right-hand panel come from 
Eq.~(\ref{eq:fit-t}) using $\tau_{GP}=10^5$ (upper curve) and $10^6$ 
(lower curve).  The $\tau_{GP}=10^5$ points were obtained by simulating 
4096 Monte Carlo photons, while the $\tau_{GP}=10^6$ points were 
obtained with 512.  All of these simulations are for continuum photons, 
with the photons being started well to the blue side of \lya\ at $\nu_{\rm 
start} = \nu_\mlya+750\,$GHz.}
\end{figure*}

The Monte Carlo simulations must be used to verify the Fokker-Planck 
estimates of (i) the colour temperature $T_c^{eff}$, and (ii) the 
spin-flip rate $\tilde S_\alpha$, which describes how rapidly the spins 
relax to the colour temperature.  Results for both of these are shown in 
Fig.~\ref{f:matrix} for $T_k=2$ and $10\,$K, and at $\tau_{GP}=10^5$ 
and $10^6$.  The agreement with the fitting formulae (Eqs.~\ref{eq:fit-s} 
and \ref{eq:fit-t}) is at the $\le 3$ per cent level.  It is especially 
remarkable that the fitting formulae perform so well at reproducing the 
correct dependence of the colour temperature on $T_s$ at low $T_k$, since 
the non-equilibrium effects on the spectral distortion must be taken into 
account and the slope of the spectrum in the Doppler cores of the 
resonances (where the validity of the Fokker-Planck equation is most
questionable) is important.

\section{Simple model for spin temperature evolution}
\label{s:model}

This section presents a simple model for the evolution of $T_s$ as a 
function of redshift.  The purpose of this model is to illustrate how much 
of a difference the improvements in the physics of the \wuf\ effect can 
make in the final result; it is not claimed that they necessarily 
represent the real universe.  Only the mean brightness temperature 
perturbation $T_b$ in the 21~cm line is calculated here for simplicity.  
While foreground synchrotron radiation probably precludes a direct 
measurement of the mean signal $T_b$ \citep{1999A&A...345..380S, 
2003MNRAS.346..871O}, it is still possible that it could be determined 
indirectly using redshift space distortions.  Specifically, on linear 
scales the $\ell=4$ moment of the power spectrum of 21~cm fluctuations, 
denoted $P_{\mu^4}(k)$ by \citet{2005ApJ...624L..65B}, is simply related 
to the mean temperature and matter power spectrum via 
$P_{\mu^4}(k)=T_b^2P_\delta(k)$.  If linear scales can be observed during 
the early stages of reionization, and the cosmological parameters are 
known well enough to estimate $P_\delta(k)$, then it may become possible 
to obtain $T_b$.

A model for $T_s$ requires a model for the evolution of the CMB 
temperature, the \lya\ flux, and the gas kinetic temperature.  Of these, 
the CMB temperature is easiest: it is
\begin{equation}
T_\gamma = T_{\gamma 0}(1+z),
\end{equation}
where $T_{\gamma 0}=2.725\,$K.  The \lya\ flux is given by
\begin{equation}
J_\alpha = \frac{(1+z)^2}{4\pi}\sum_{n=2}^\infty P_{np}
\int_z^{z_{\rm max}} \frac{c}{H(z')}\epsilon(\nu'_n,z')
\,\rmd z';
\end{equation}
see \citet{2004astro.ph.10129B}.  The UV source term is 
$\epsilon(\nu'_n,z')$, which is the number of photons emitted per unit 
comoving volume per unit proper time per unit frequency at redshift $z'$ 
and frequency $\nu'_n$ (see below).
The factor of $P_{np}$ has been added to 
account for the fact that not all photons in the 912--1216$\,$\AA\ 
band degrade to \lya.  The 
$n$th term in the sum is the contribution from photons emitted between the 
$1s\rightarrow np$ and $1s\rightarrow (n+1)p$ Lyman transitions, which 
ultimately redshift and excite $1s\rightarrow np$; as such, the emitted 
photon frequency is $\nu'_n = \nu_{1s\rightarrow np}(1+z')/(1+z)$
and the maximum redshift from which this photon could have been received 
is
\begin{equation}
1+z_{\rm max} = \frac{\nu_{1s\rightarrow n+1,p}
}{\nu_{1s\rightarrow np}}(1+z)
= \frac{1-(n+1)^{-2}}{1-n^{-2}}(1+z).
\end{equation}

The source emissivity is modeled following \citet{2004astro.ph.10129B} by 
the equation
\begin{equation}
\epsilon(\nu,z) = \epsilon_b(\nu) \frac{\Omega_b}{m_p\Omega_m}
\frac{\rmd}{\rmd t}\int f_\star(M,t) Mn(M,t) \rmd M,
\label{eq:epsuv}
\end{equation}
where $M$ is the relevant halo mass, $n(M,t)$ is the comoving number 
density of halos at proper time $t$ per unit mass, $f_\star(M,t)$ is the 
fraction of the 
baryons that have turned into stars, and $\epsilon_b(\nu)$ is the number 
of photons emitted per baryon by the stars.  This equation assumes that 
the lifetimes of the UV-emitting stars are short compared to the Hubble 
time, so that the UV emissivity tracks the instantaneous star formation 
rate.  This is reasonable since most radiation at 912--1216$\,$\AA\ 
is emitted by the most massive stars with lifetimes of 
$<10^7\,$years, whereas the Hubble time during reionization is
$>10^8\,$years.  The halo mass function of \citet{1999MNRAS.308..119S} is 
used.

We consider the temperature evolution of the IGM due to cosmological 
expansion and X-ray heating.  The real universe has inhomogeneities that 
alter the spin temperature evolution via changes in the kinetic 
temperature (in shocks, by adiabatic expansion or compression during 
structure formation, or from inhomogeneous X-ray sources), and by 
enhancing the collisional coupling in the denser regions.  The main effect 
is to increase the 21~cm emissivity of halos and filaments 
\citep{2002ApJ...572L.123I, 2005astro.ph..9651A, 2005astro.ph.10814K} and 
so including them in the model would make the computed signal more 
positive (or less negative).  For example, \citet{2005astro.ph..9651A} 
find in a simulation with no X-ray or UV sources that these effects 
increase the mean signal by $+1\,$mK at $z=18$ and $+5\,$mK at $z=10$.  We 
have not simulated the effect of inhomogeneities in the presence of
UV radiation, but they could be larger than found by
\citet{2005astro.ph..9651A} because the \wuf\ coupling will make most of 
the diffuse, unshocked IGM ``visible'' and hence the importance of 
temperature fluctuations in the unshocked phase will be increased.
Subject to these caveats, our temperature evolution equation is thus
\begin{equation}
(1+z)\frac{\rmd T_k}{\rmd z} = 2T_k - 
\frac{2\mu m_p\Gamma_X}{3\rho_{b0}k_BH(z)},
\label{eq:thermal}
\end{equation}
assuming a monatomic gas \citep{2004ApJ...602....1C} with mean atomic
weight of $\mu=1.22$, as appropriate for a hydrogen-helium mixture with 
helium mass fraction 0.24.

The X-ray heating $\Gamma_X$ (in, e.g. ergs per physical second per 
comoving cm$^3$) is
\begin{equation}
\Gamma_X = f_\Gamma f_{Xe} E_X \frac{\Omega_b}{m_p\Omega_m}
\frac{\rmd}{\rmd t}\int f_\star(M,t) Mn(M,t) \rmd M,
\label{eq:gx}
\end{equation}
where $f_\Gamma$ is the fraction of X-ray energy that goes into heating 
the IGM, $f_{Xe}$ is the fraction of X-ray photons that escape from an 
early star cluster or galaxy,
and $E_X$ is the energy emitted in X-rays per baryon that forms 
stars.  There are many sources that contribute to $E_X$, e.g. 
stars, supernovae, X-ray binaries, and quasars, and both the total X-ray 
emission and the relative contributions from different sources are very 
uncertain \citep{2003MNRAS.340..210G}.  Also Eq.~(\ref{eq:gx}) assumes 
that the X-ray heating tracks the star formation rate, which may not be 
true particularly if quasars contribute significantly to the X-ray 
emission.  Equation~(\ref{eq:thermal}) has the solution
\begin{eqnarray}
T_k(z) &=& \left(\frac{1+z}{1+z_0}\right)^2T_k(z_0)
\nonumber \\ &&
+ \frac{2\mu m_p}{3\rho_{b0}k_B}\int_z^{z_0}
\frac{\Gamma_X(z')}{H(z')} \frac{(1+z)^2}{(1+z')^3}\rmd z',
\end{eqnarray}
where $z_0$ is an arbitrary starting redshift, which can be any time 
after the thermal decoupling of the gas from the CMB but before heating is 
important.  We use $z_0=50$ and initialize the temperature using {\sc 
recfast} \citep{1999ApJ...523L...1S}.

\begin{figure}
\includegraphics[angle=-90,width=3.2in]{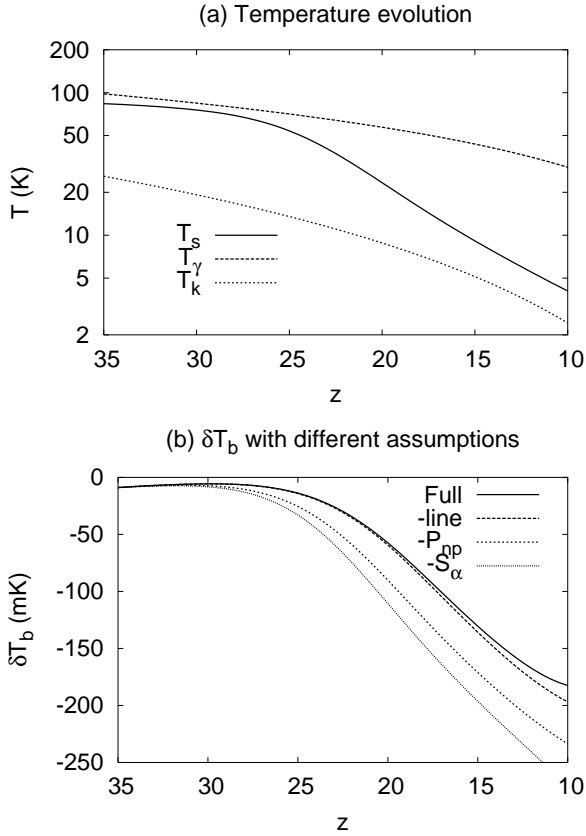}
\caption{\label{f:anx}(a) Spin temperature evolution assuming Population 
III stars forming with efficiency $f_\ast=2.5\times 10^{-4}$ in haloes 
that can cool via atomic transitions.  All sources of heating, including 
X-rays, have been neglected in this model.
(b) The effect of changing the physics of the \wuf\ effect.  The solid 
curve shows the full calculation for the mean brightness temperature 
$T_b$.  The long-dashed curve shows the 
calculation removing the spin diffusivity and fine structure corrections.  
The short-dashed curve also assumes $P_{np}=1$ instead of the correct 
values; this is the curve that would be calculated using the most recent 
models prior to this paper.  The dotted curve makes the further 
simplification that $S_\alpha=1$, as was done by 
\citet{1997ApJ...475..429M}.}
\end{figure}

\begin{figure}
\includegraphics[angle=-90,width=3.2in]{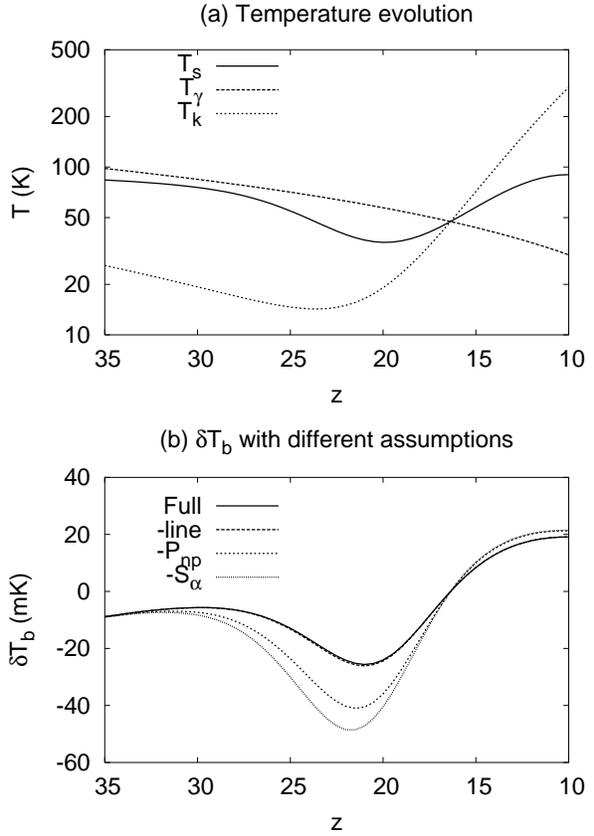}
\caption{\label{f:awx}Same as Fig.~\ref{f:anx}, except with X-rays.  Here 
it is assumed that the X-rays escaping from early galaxies carry 
0.5 per cent of the stellar energy output, corresponding to 
$f_{Xe}E_X=27\,$keV.}
\end{figure}

An example of this model is shown in Fig.~\ref{f:anx}.  Here it is assumed 
that stars form only in haloes with virial temperature $T_{\rm 
vir}>10^4\,$K that can cool via atomic transitions.  The star formation 
efficiency is taken as $f_\ast=2.5\times 10^{-4}$ in these haloes, which 
causes the 
\lya\ coupling to turn on ($x_\alpha=1$) at $z\approx 21$.  Their 
$\epsilon_b(\nu)$ is assumed to be a blackbody of temperature $10^5\,$K, 
as appropriate for massive Population III stars with $M\ge 300M_\odot$ 
\citep{2001ApJ...552..464B}; the blackbody is normalized to a total energy 
of 7.1 MeV per H nucleus or 5.4 MeV per baryon, appropriate for complete 
hydrogen burning to $^4$He.  (Most of the star's energy is released during 
the hydrogen-burning stage.)  This model contains no X-ray emission.  If 
one assumes that 0.5 per cent of the stars' energy emerges from early 
galaxies in the form of X-rays that can heat the IGM (corresponding to 
$f_{Xe}E_X=27\,$keV), and that the heating
efficiency is $f_\Gamma=0.14$ \citep{1985ApJ...298..268S, 
2004ApJ...602....1C}, then one obtains the model in 
Fig.~\ref{f:awx}.\footnote{This amount of X-ray emission corresponds to 
$\alpha_X=0.028$ in the notation of \citet{2004ApJ...602....1C}.}  In both 
cases, the best-fit 6-parameter cosmology of \citet{2004astro.ph..7372S} 
was used.

In both the examples with and without X-ray emission, a calculation 
neglecting the \lya\ spectral distortion (e.g. 
\citealt{1997ApJ...475..429M}) can overestimate the 21~cm signal by as 
much as a factor of $\sim 2.4$, as shown by the dotted curves. 
Incorporating the simplified model of the \lya\ spectral distortion using 
the Voigt profile (e.g. \citealt{2004ApJ...602....1C}) reduces the error 
to a factor of $\le 1.9$, as shown by the short-dashed curves.  Most of 
the remaining error is due to the two-photon decays (included in the 
long-dashed curves). The inclusion of \lya\ fine structure and spin 
diffusivity (solid line) makes a $<10$ per cent difference in the model 
with no X-rays and even less in the model with X-rays.  Thus it is seen 
that the two-photon correction $P_{np}$ can have a large effect on the 
21~cm signal.

\section{Conclusions}
\label{s:conc}

The \hi\ spin temperature of the IGM is determined by a balance of 
interaction with the CMB in the 21~cm line, atomic collisions, and the 
\wuf\ effect.  The last of these depends on both the emission rate of UV 
photons and on the coupling coefficient $P_{np}S_\alpha$.  In this paper, 
I have evaluated the coupling coefficient including several new physical 
processes, and found that it is lower than previously computed.  The most 
important correction is the inclusion of two-photon decay, $P_{np}<1$.  
Fine and hyperfine structure effects and spin diffusivity are small except 
at low temperatures.  The Fokker-Planck equation is found to provide an 
accurate description of the \wuf\ effect at the several per cent level 
even at the lowest temperatures that could reasonably be encountered in 
the IGM.  Fitting formulae for the scattering rate $\tilde S_\alpha$ 
(Eq.~\ref{eq:fit-s}) and colour temperature $T_c^{eff}$ 
(Eq.~\ref{eq:fit-t}) have been provided, along with bounds on their 
errors.

The corrections described here pertain to the strength of the \wuf\ effect 
and are important only during the era when $x_c<x_\alpha\le O(1)$.  Early 
on ($z>30$ in the models of \S\ref{s:model}), the \wuf\ effect is 
negligible.  Later on ($z<15$ in the models of \S\ref{s:model}), the \wuf\ 
effect becomes saturated in the sense that $x_\alpha\gg 1$ and $T_s\approx 
T_k$; in this case changes in the coupling strength have no impact on the 
observable temperature fluctuations.  The changes described here, 
particularly $P_{np}$, can however have a very large effect at 
intermediate redshifts (here $15<z<30$) particularly where $x_\alpha\sim 
1$.  This is the range of redshifts at which \citet{2004astro.ph.10129B} 
have suggested that the fluctuations in the \lya\ background could be 
observable, providing information about early galaxies such as their bias 
(and hence their halo mass).  These authors found that photons redshifting 
into the higher-order Lyman transitions Ly$n$ ($n\gg 1$) dominate the 
\lya\ fluctuations at $k\ge 0.1h\,$Mpc$^{-1}$; since $P_{np}=0.36$ for 
these photons, the power spectrum of these small-scale \lya\ fluctuations 
will be correspondingly reduced.  For this application in particular, the 
inclusion of the two-photon decay mechanism will be valuable in extracting 
maximal information from 21~cm observations.

\section*{Acknowledgments}

I wish to thank P. J. E. Peebles, Uro{\v s} Seljak, and Kris Sigurdson for 
useful comments, and X. Chen and J. Miralda-Escud\'{e} for discussion 
about the high-temperature behaviour of the spectral distortion.  I 
acknowledge support from NASA grant NGT5-50383.

\appendix

\section{The fate of hydrogen in the \twos\ configuration}
\label{s:2photon}

When excited hydrogen atoms decay to the $2s$ configuration, there are no 
further one-photon transitions allowed to the ground configuration.  In 
vacuum, the $2s$ configuration decays via two-photon emission with a rate 
$\Lambda=8.2\,$s$^{-1}$ \citep{goeppert1931, 1940ApJ....91..215B, 
1959PhRv..113..179S}.  The purpose of this appendix is to show that under 
conditions encountered in the high-redshift IGM, the two-photon process is 
faster than competing processes, namely collisions and interactions with 
the CMB.

The $2s_{1/2}$ level can be depopulated by either collisions with neutral 
atoms or with charged particles.  Only crude approximations to 
these are needed since they will be shown to be negligible.  This 
is convenient, 
since there are no published rate coefficients for the low temperatures 
required here. For the collisions with neutral H or He, the rate of 
de-excitation of $2s_{1/2}$ is $\Gamma_Q\sim n\sigma_Q v$, where $n$ is 
the number density of H or He, $\sigma_Q$ is the quenching cross section, 
and $v$ is the typical velocity $v\sim 1.3\times 
10^4T_k^{1/2}\,$cm$\,$s$^{-1}$ (with $T_k$ in Kelvins).  In order for the 
collisional de-excitation rate to be 1 per cent of the two-photon rate at 
$z=75$, one needs $\sigma_Q \sim 10^{-4}T_k^{-1/2}\,$cm$^2$, i.e. many 
orders of magnitude larger than the cross sections for collision of 
neutral atoms; thus the atomic collisions contribute negligibly to the 
de-population of H($2s_{1/2}$). Of course in the standard cosmological 
model there are no \lya\ photons at $z=75$; the result that collisions 
with neutrals are negligible is even stronger at lower redshifts that are 
more reasonable for the \wuf\ effect.

Cross sections for H$(2s_{1/2})$ with charged particles (e$^-$ or p$^+$) 
can be much larger than for neutral atoms, particularly at low 
temperature, because the long-range electric field of the passing charged 
particle can produce a Stark effect that mixes $2s$ and $2p$; once the H 
atom reaches $2p$, it decays quickly by \lya\ emission.  The rate 
coefficients $W$ (in e.g. cm$^3\,$s$^{-1}$) scale roughly as $T_k^{-1/2}$ 
and are dominated by collision with protons \citep{1955PPSA...68..457S}.  
Extrapolating the rate coefficients from \citet{1955PPSA...68..457S} at 
$T_k=10^4\,$K down, and assuming no heating of the IGM so that 
$T_k=0.022(1+z)^2\,$K, one finds
a rate coefficient of $W=0.36(1+z)^{-1}\,$cm$^3\,$s$^{-1}$.  This is an 
upper limit because the actual scaling is shallower than $W\propto 
T_k^{-1/2}$ at
low $T_k$, and because any heating of the IGM increases $T_k$.
The rate of charged particle collisional de-excitation is then
\begin{equation}
\Gamma_{\rm charge}\sim \nh x_e W = 7\times 10^{-4}\left(
\frac{1+z}{100}\right)^{2}x_e \,{\rm s}^{-1},
\end{equation}
which is much less than the two-photon rate $\Lambda=8.2\,$s$^{-1}$ 
at all relevant redshifts since the electron-to-hydrogen nucleus ratio 
$x_e$ is always less than 1.16 (and much less before reionization).

The CMB can depopulate the \hi\ $2s_{1/2}$ level via stimulated emission 
at the Lamb shift frequency $\nu_{1/2} = 1.06\,$GHz to the $2p_{1/2}$ 
level, or via radiative excitation to $2p_{3/2}$ at 
$\nu_{3/2}=11\,$GHz.  \hi\ atoms in these levels decay by \lya\ emission.
The rates for these are given by the usual formula
\begin{equation}
\Gamma(2s_{1/2}\rightarrow 2p_j) = \frac{32\pi^3\alpha\nu_j^3}{3c^2}
\overline{\sum|{\bmath r}_{2p_j,2s_{1/2}}|^2}
\left(\frac{k_BT_\gamma}{h\nu_j}\right),
\label{eq:g2s2p}
\end{equation}
where the bar and summation indicates that the squares of the dipole 
matrix elements 
${\bmath r}_{2p_j,2s_{1/2}}$ are averaged over 
values of the magnetic quantum number in the $2s_{1/2}$ level and summed 
over the $2p_j$ level, and
the last factor is the number of photons per state.  (This is much greater 
than 1 so spontaneous emission and quantum corrections to the 
Rayleigh-Jeans formula can be neglected.)  The dipole matrix elements
$\overline{\sum|{\bmath r}_{2p_j,2s_{1/2}}|^2}$ are $9a_0^2$ for 
$j=\frac{1}{2}$ and $18a_0^2$ for $j=\frac{3}{2}$, where $a_0$ is the Bohr 
radius.  Substituting into Eq.~(\ref{eq:g2s2p}) and using 
$T_\gamma=2.73(1+z)\,$K yields
\begin{eqnarray}
\Gamma(2s_{1/2}\rightarrow 2p_{1/2}) &=& 4.4\times 10^{-8}
(1+z)\,{\rm s}^{-1}
{\rm ~~and}
\nonumber \\
\Gamma(2s_{1/2}\rightarrow 2p_{3/2}) &=& 9.3\times 10^{-6}
(1+z)\,{\rm s}^{-1}.
\end{eqnarray}
These rates are negligible compared with the 2-photon rate 
$\Lambda=8.2\,$s$^{-1}$ at all relevant redshifts.

Most of the CMB photons during the reionization era have much higher 
energies than 1.06 or 11 GHz (for comparison, $k_BT_\gamma/h=570\,$GHz at 
$1+z=10$).  These photons can cause nonresonant Raman scattering, 
$2s_{1/2}\rightarrow 1s_{1/2}$, that puts the hydrogen atom in the ground 
state and results in the emission of a photon with frequency just above 
the \lya\ frequency.  This photon immediately redshifts into the \lya\ 
doublet and can participate in the \wuf\ effect.  The relevant frequencies 
are all much greater than the fine structure splitting, so at least for a 
rough estimate one can ignore electron spin in the calculation of the 
Raman scattering rate.  The Raman scattering cross section is (e.g. 
\citealt{1971rqt..book.....B}, Eq.~61.8)
\begin{eqnarray}
\sigma(i\rightarrow f) \!\! &=&
\frac{128\pi^5e^4\nu\nu'{^3}}{9h^2c^4}
\nonumber \\ && \times
\sum_{\alpha,\beta}
\left|\sum_{m_l=-1}^1\frac{\langle 1s|r^\alpha|2p,m_l\rangle
\langle 2p,m_l|r^\beta|2s\rangle}{\Delta\nu}
\right|^2
\nonumber \\
&=& \frac{128\pi^5e^4\nu\nu'{^3}}{27h^2c^4}
\left|\frac{\langle 1s||{\bmath r}||2p\rangle
\langle 2p||{\bmath r}||2s\rangle}{\Delta\nu}
\right|^2,
\end{eqnarray}
where $\nu$ is the incoming frequency, $\nu'=\nu_\mlya+\nu$ is the 
outgoing frequency, and $\Delta\nu$ is 
the detuning from the intermediate ($2p$) state, i.e.
\begin{equation}
h\Delta\nu_n = h\nu + E_{2s}- E_{2p}.
\end{equation}
This includes only the $2p$ intermediate state since the total energy 
of the atom and photon is only slightly above the $n=2$
energy level, hence its denominator $\Delta\nu$ is the largest.  For the 
same reason the terms in the Raman matrix element where 
the outgoing photon is emitted before the incoming photon is absorbed 
have been dropped.
One also has $\nu'\approx \nu_\mlya$, and because of the $2s$--$2p$ 
degeneracy $\Delta\nu_{2p} \approx \nu$.  Putting this together and using 
the hydrogenic matrix elements gives
\begin{equation}
\sigma(i\rightarrow f) = \frac{4194304\pi^5e^4\nu_\mlya^3a_0^4}
{19683h^2c^4\nu}.
\end{equation}
The total Raman scattering rate (per atom in the $2s_{1/2}$ level) is 
given by integration of the cross section over the blackbody curve:
\begin{eqnarray}
\Gamma_{\rm Raman} &=& \int \frac{8\pi\nu^2}{c^2
(\rme^{h\nu/k_BT_\gamma}
-1) } \sigma(i\rightarrow f) \rmd\nu
\nonumber \\
&=& \frac{16777216\pi^8e^4\nu_\mlya^3a_0^4k_B^2}
{59049h^4c^6}T_\gamma^2
\nonumber \\
&=& \frac{64\pi^2\alpha^6k_B^2T_\gamma^2}{81 h^2\nu_\mlya}
= 1.7\times 10^{-7}(1+z)^2\,{\rm s}^{-1},
\end{eqnarray}
where we have used $a_0=3e^2/(8h\nu_\mlya)$ and 
$\alpha = 2\pi e^2/(hc)$.  Once again, this rate is negligible compared to 
$\Lambda=8.2\,$s$^{-1}$ at the redshifts of interest for the \wuf\ effect.

% The 1s2p and 2s2p reduced matrix elements are 2^15/3^9 a_0^2
% and 27 a_0^2 respectively.

\section{\lya\ cross section}
\label{s:lya}

In order to compute the \wuf\ coefficient $x_\alpha$, it is necessary to 
know the cross sections for resonant Rayleigh and Raman scattering between 
the two hyperfine levels $1s_{1/2}(F=0,1)$.  There are four cross sections 
$F\rightarrow F'$, where $F,F'\in\{0,1\}$, which depend on the photon 
frequency $\nu$.  Similar computations can be found in 
\citet{1988ApJ...334..527D} and \citet{1998A&A...332..732B}, but this 
Appendix includes both the hyperfine structure and the detailed frequency 
dependence.

The cross sections can be determined from the reduced dipole matrix 
elements between the $1s_{1/2}(F)$ and $2p_j(F')$ hyperfine levels.  The 
electron position operator ${\bmath r}$ has reduced matrix element given 
by the hydrogenic form
\begin{equation}
\langle 2p || {\bmath r} || 1s \rangle = \frac{128\sqrt{6}}{243}a_0.
\end{equation}
Since the ${\bmath r}$ operator acts only on the electron's positional 
degrees of freedom, without regard to electronic or nuclear spin, the 
hyperfine matrix elements can be obtained entirely from group theory.  
Applying Eq.~(7.1.7) of \citet{1960amqm.book.....E} twice, and
using the fact that \hi\ has electronic spin $S=\frac{1}{2}$
and (for $^1$H) nuclear spin $I=\frac{1}{2}$,
\begin{equation}
\langle n'l'{_{j'}}(F') || {\bmath r} || nl_j(F) \rangle
= {\cal I}\langle n'l' || {\bmath r} || nl \rangle,
\label{eq:k}
\end{equation}
where the coefficient ${\cal I}$ is given by the $6j$ symbols,
\begin{eqnarray}
{\cal I} &=& (-1)^{l'+j+j'+F+1}
\nonumber \\ && \times
[(2j+1)(2j'+1)(2F+1)(2F'+1)]^{1/2}
\nonumber \\ && \times
\sixj{l'}{j'}{\frac{1}{2}}{j}{l}{1}
\sixj{j'}{F'}{\frac{1}{2}}{F}{j}{1}.
\end{eqnarray}
Values of ${\cal I}$ are shown in Table~\ref{t:lya}.

The matrix element for resonant electric dipole scattering with incoming 
photon energy $h\nu$ is
\begin{equation}
(c^{\mu\nu})_{fi} = e^2 \sum_a
  \frac{\langle f|\hat r^\mu|a\rangle\langle a|\hat r^\nu|i\rangle}
  {E_a - E_i - h\nu - ih\Gamma_a/4\pi},
\end{equation}
where $\Gamma_a$ is the width of the intermediate state $a$ and the width 
of the initial and final states is neglected.  In order to obtain the 
differential cross section for $F_i\rightarrow F_f$ scattering by 
randomly oriented atoms, one must obtain the spin-$K$ irreducible parts of 
the scattering tensor,
\begin{equation}
\bar G_{F_i\rightarrow F_f}^{(K)} = 
\frac{\Pi^{(K)}_{\mu\nu\alpha\beta}}{2F_i+1} \sum_{M_f=-F_f}^{F_f}
\sum_{M_i=-F_i}^{F_i}
(c^{\mu\nu})_{fi} (c^{\alpha\beta})_{fi}^\ast,
\label{eq:gbar}
\end{equation}
where $\Pi^{(K)}_{\mu\nu\alpha\beta}$ is the projection matrix that 
selects the spin-$K$ ($K=0,1,2$) part of an arbitrary second-rank tensor 
$X_{\alpha\beta}$.  This decomposition of second-rank tensors is complete, 
so that
\begin{equation}
\sum_{K=0}^2 \Pi^{(K)}_{\mu\nu\alpha\beta} = g_{\mu\alpha}
g_{\nu\beta},
\end{equation}
where $g_{\mu\nu}$ is the metric tensor, equal to $\delta_{\mu\nu}$ in the 
usual Cartesian coordinate basis.  The most convenient basis for these 
calculations, however, is not the Cartesian basis but the polar basis 
\citep{1960amqm.book.....E} in which the coordinates $r^\mu$ are related 
to Cartesian $X$, $Y$, and $Z$ via
\begin{equation}
r^{\pm 1} = \mp\frac{1}{\sqrt{2}}(X\pm iY) {\rm ~and~}
r^0 = Z;
\end{equation}
the metric tensor is $g_{\mu\nu}=(-1)^\mu\delta_{\mu,-\nu}$.  
In this basis the powerful spherical tensor methods can be used.  The 
projection matrix is then
\begin{eqnarray}
\Pi^{(K)}_{\mu\nu\alpha\beta} &=& \sum_{M_K,\gamma,\delta} 
g_{\mu\gamma}g_{\nu\delta}
\langle 1\gamma;1\delta|KM_K\rangle
\nonumber \\ && \times
\langle KM_K|1\alpha;1\beta\rangle
\nonumber \\
&=& (2K+1) \sum_{M_K=-K}^K (-1)^{M_K}
\threej{1}{1}{K}{\mu}{\nu}{-M_K}
\nonumber \\ && \times
\threej{1}{1}{K}{\alpha}{\beta}{M_K},
\label{eq:pih}
\end{eqnarray}
where in the first line Clebsch-Gordon coefficients have been used to 
emphasize the nature of $\Pi^{(K)}$ as a projection matrix, and in 
the second line these have been converted to $3j$ symbols.

Writing Eq.~(\ref{eq:gbar}) in terms of reduced matrix elements, and 
substitutes Eq.~(\ref{eq:pih}), one obtains
\begin{eqnarray}
\bar{G}^{(K)}_{F_i\rightarrow F_f} \!\!\! &=&
\frac{2K+1}{2F_i+1} \frac{e^4}{h^2}\sum_{M_f,M_i,M_K,\mu,\nu,\alpha,\beta
  ,a,b}
\nonumber \\ &&
\frac{\langle f || {\bmath r} || a\rangle
\langle a || {\bmath r} || i\rangle \langle i || {\bmath r} || b\rangle
\langle b || {\bmath r} || f\rangle}
  {(\Delta\nu_{ai}+i\frac{\Gamma_a}{4\pi})
   (\Delta\nu_{bi}-i\frac{\Gamma_b}{4\pi})}
  (-1)^{M_K}
\nonumber \\ && \times
  \threej{1}{1}{K}{\mu}{\nu}{-M_K}
  \threej{1}{1}{K}{\alpha}{\beta}{M_K}
\nonumber \\ && \times
  (-1)^{F_f-M_f}\threej{F_f}{1}{F_a}{-M_f}{\mu   }{M_a}
\nonumber \\ && \times
  (-1)^{F_a-M_a}\threej{F_a}{1}{F_i}{-M_a}{\nu   }{M_i}
\nonumber \\ && \times
  (-1)^{F_i-M_i}\threej{F_i}{1}{F_b}{-M_i}{\alpha}{M_b}
\nonumber \\ && \times
  (-1)^{F_b-M_b}\threej{F_b}{1}{F_f}{-M_b}{\beta }{M_f}.
\end{eqnarray}
The complicated sums of $3j$ symbols can be reduced by applying the 
reduction formula (Eq. 6.2.8 of \citealt{1960amqm.book.....E}) twice and 
then using $3j$ symbol orthogonality.  This eliminates all summation over 
magnetic quantum numbers:
\begin{eqnarray}
\bar{G}^{(K)}_{F_i\rightarrow F_f} \!\!\! &=& 
\frac{2K+1}{2F_i+1}\frac{e^4}{h^2} \sum_{a,b} (-1)^{F_a-F_b}
\nonumber \\ && \times
\frac{\langle f || {\bmath r} || a\rangle
\langle a || {\bmath r} || i\rangle \langle i || {\bmath r} || b\rangle
\langle b || {\bmath r} || f\rangle}
  {(\Delta\nu_{ai}+i\frac{\Gamma_a}{4\pi})
   (\Delta\nu_{bi}-i\frac{\Gamma_b}{4\pi})}
\nonumber \\ && \times
\sixj{F_f}{F_i}{K}{1}{1}{F_a}
\sixj{F_f}{F_i}{K}{1}{1}{F_b}.
\label{eq:gbar-6j}
\end{eqnarray}
Similar expressions are given by \citet{1972ApJ...175..185O} and 
\citet{1988ApJ...334..527D} for the case where there is a single (possibly 
degenerate) intermediate level.

The cross section is given by Eqs.~(61.7) and (61.9) of 
\citet{1971rqt..book.....B}\footnote{\citet{1971rqt..book.....B} denote 
our $\bar G^{(0)}$, $\bar G^{(1)}$, and $\bar G^{(2)}$ by $3G^0$, $G^a$, 
and $G^s$, respectively.  This can be seen by noting that 
their Eq.~(60.10) is precisely the projection 
$\Pi^{(K)}$, $K=0,1,2$, but in the Cartesian instead of the polar basis.}.  
Noting that the phase space factors involving the frequency can be 
evaluated at $\nu_\mlya$ with negligible error yields a total cross 
section
\begin{equation}
\sigma_{F_i\rightarrow F_f} = \frac{128\pi^5\nu_\mlya^4}{9c^4}(\bar 
G^{(0)}  + \bar G^{(1)} + \bar G^{(2)}).
\end{equation}
The angular dependence is given by
\begin{equation}
\frac{\rmd\sigma_{F_i\rightarrow F_f}}{\rmd\Omega} = 
\frac{\sigma_{F_i\rightarrow F_f}}{4\pi}[1 + 
5\varpi_{2;F_i\rightarrow F_f}P_2(\cos\theta)],
\label{eq:varpi}
\end{equation}
where $P_2$ is a Legendre polynomial and the phase function is
\begin{equation}
\varpi_{2;F_i\rightarrow F_f} = \frac{\frac{1}{10}\bar G^{(0)} - 
\frac{1}{20}\bar G^{(1)}
  + \frac{1}{100}\bar G^{(2)}}
{\bar G^{(0)} + \bar G^{(1)} + \bar G^{(2)}}.
\end{equation}
Repeated scattering of \lya\ photons eliminates any polarization so the 
polarization dependence is not needed.

\begin{table}
\caption{\label{t:lya}The six hyperfine components of the \lya\ lines.  
The frequency offset shown is relative to the lowest-frequency line, i.e. 
it is $\nu-\nu_A$.  The values of ${\cal I}$ are used in 
Eq.~(\ref{eq:k}).}
\begin{tabular}{ccccr}
\hline\hline
Line & Lower & Upper & ${\cal I}$ & $\!\!\!\!$Frequency \\
 & level & level & & offset \\
 & & & & (GHz) \\
\hline
A & $1s_{1/2}(F=1)$ & $2p_{1/2}(F=0)$ & $+\sqrt{1/3}$ &  0.000 \\
B & $1s_{1/2}(F=1)$ & $2p_{1/2}(F=1)$ & $-\sqrt{2/3}$ &  0.059 \\
C & $1s_{1/2}(F=0)$ & $2p_{1/2}(F=1)$ & $-\sqrt{1/3}$ &  1.479 \\
D & $1s_{1/2}(F=1)$ & $2p_{3/2}(F=1)$ & $-\sqrt{1/3}$ & 10.945 \\
E & $1s_{1/2}(F=1)$ & $2p_{3/2}(F=2)$ & $+\sqrt{5/3}$ & 10.968 \\
F & $1s_{1/2}(F=0)$ & $2p_{3/2}(F=1)$ & $+\sqrt{2/3}$ & 12.365 \\
\hline\hline
\end{tabular}
\end{table}

The scattering cross-sections can then be determined in terms of the 
detunings for the six hyperfine transitions of \lya, shown in 
Table~\ref{t:lya}, and their half-width at half-maximum (HWHM) widths,
\begin{equation}
\gamma=\frac{\Gamma_{2p}}{4\pi}
= \frac{16\pi^3e^2\nu_\mlya^3}{9hc^3}\left| \langle 2p || {\bmath r} ||
1s \rangle \right|^2 =50\,{\rm MHz}
\label{eq:gamma}
\end{equation}
($\gamma$ is the same for all $2p$ levels on account of the sum rules).  
In writing the cross sections, it is convenient to define the normalized 
Lorentzian profiles
\begin{equation}
\phi_{AA} = \frac{\gamma}{\pi(\Delta\nu_A^2+\gamma^2)}
\end{equation}
and the interference profiles
\begin{equation}
\phi_{AB} = \frac{\gamma(\Delta\nu_A\Delta\nu_B+\gamma^2)
}{\pi(\Delta\nu_A^2+\gamma^2)(\Delta\nu_B^2+\gamma^2)}.
\end{equation}
(Similar definitions are used for profiles of other lines.)
The cross-sections in the atom's rest frame are
\begin{equation}
\sigma(F_i\rightarrow F_f) = \frac{3}{8\pi}\lambda_\mlya^2\Gamma_{2p}
\phi^u_{F_iF_f},
\label{eq:sigphi}
\end{equation}
where
\begin{eqnarray}
\phi^u_{00} &=& 
\frac{1}{9}\phi_{CC}+\frac{4}{9}\phi_{FF}+\frac{4}{9}\phi_{CF},
\nonumber \\
\phi^u_{11} &=& \frac{1}{9}\phi_{AA} + \frac{4}{27}\phi_{BB}
  + \frac{1}{27}\phi_{DD} + \frac{5}{9}\phi_{EE} + \frac{4}{27}\phi_{BD},
\nonumber \\
\phi^u_{01} &=& 
\frac{2}{9}\phi_{CC}+\frac{2}{9}\phi_{FF}-\frac{4}{9}\phi_{CF},
{\rm ~and}
\nonumber \\
\phi^u_{10} &=& \frac{2}{27}\phi_{BB} + \frac{2}{27}\phi_{DD}
  -\frac{4}{27}\phi_{BD}.
\nonumber \\
\label{eq:phiff}
\end{eqnarray}
(The $^u$ superscript indicates that these profiles are un-convolved and 
do not include thermal broadening.)
These satisfy the line profile normalization conditions
\begin{equation}
\sum_{F_f=0}^1\int_{-\infty}^\infty 
\phi^u_{F_iF_f}(\Delta\nu)\;\rmd\Delta\nu = 1.
\end{equation}
In gas with a finite temperature, all profiles must be 
convolved with a Gaussian of $1\sigma$ width
\begin{equation}
\sigma_\nu = \sqrt{\frac{k_BT}{m_pc^2}}\, \nu_\mlya,
\label{eq:signu}
\end{equation}
that is,
\begin{equation}
\phi_{F_iF_f}(\nu) = \int_{-\infty}^\infty
\phi^u_{F_iF_f}(\nu') \frac{1}{\sqrt{2\pi}\,\sigma_\nu}
\rme^{-(\nu-\nu')^2/2\sigma_\nu^2}
\,\rmd\nu'.
\end{equation}

The phase factors $\varpi_{2;F_i\rightarrow F_f}$ are
\begin{eqnarray}
\varpi_{2;0\rightarrow 0} \!\!\! &=& \frac{1}{10},
\nonumber \\
\varpi_{2;1\rightarrow 1} \!\!\! &=& \frac{1}{40}(
4\phi_{BB}+\phi_{DD}+21\phi_{EE}+24\phi_{AE}
\nonumber \\ && \;\;\;\;
  + 4\phi_{BD} + 36\phi_{BE} + 18\phi_{DE})
\nonumber \\ && \times
(\phi_{AA}+4\phi_{BB}+\phi_{DD}
+15\phi_{EE}+4\phi_{BD})^{-1},
\nonumber \\
\varpi_{2;0\rightarrow 1} \!\!\! &=& -\frac{1}{20},
{\rm ~and}\nonumber \\
\varpi_{2;1\rightarrow 0} \!\!\! &=& -\frac{1}{20}.
\label{eq:varpi-eval}
\end{eqnarray}
Note that $\varpi_{2;1\rightarrow 1}$ is frequency-dependent because there 
are several resonances with different symmetries that contribute to it.  
This frequency must of course be evaluated in the atom frame rather than 
the frame at rest with respect to the bulk gas.

\section{Random velocity generator}
\label{app:rng}

This Appendix presents an algorithm for generating random variables 
$u_\parallel$ from the distribution of Eq.~(\ref{eq:pu}).  This 
distribution is an appropriately normalized version of a Gaussian (the 
Maxwellian velocity distribution of the H atoms) times a resonance line 
profile. In this case the resonance line profile is complicated and has up 
to four separate resonances, including interference terms.  There are 
existing algorithms \citep{1977ApJ...218..857L, 1982ApJ...255..303L} for 
the case where the resonance line profile is Lorentzian in the atom frame, 
and the algorithm given here draws on many of the same concepts.  Our 
version of the algorithm is not highly optimized and there are places 
where it could be sped up significantly at the expense of additional 
complexity, but its speed is adequate for our purposes.  In particular, 
the code is fast within 1--2$\sigma_\nu$ of the Doppler cores of the \hi\ 
$1s_{1/2}$--$2p_{1/2}$ and $1s_{1/2}$--$2p_{3/2}$ lines, and since nearly 
all scatterings occur in these regions there is little to be gained by 
speeding up the code at other frequencies.

\begin{figure}
\includegraphics[width=2.6in]{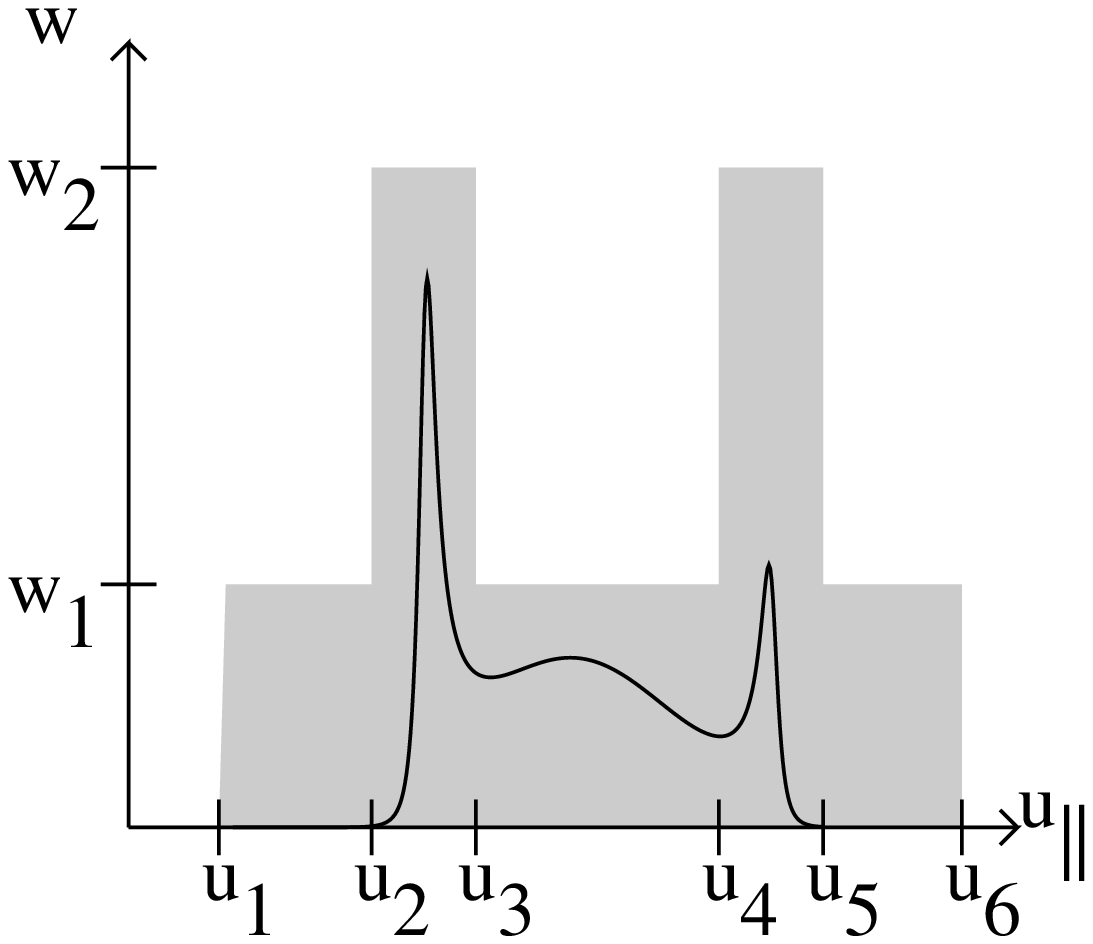}
\caption{\label{f:rect}The region chosen for the rejection algorithm.  The 
shaded region corresponds to the region within which $(u_\parallel,w)$ 
pairs are chosen, and the solid line indicates the upper boundary of the 
acceptance region.  The scale in the figure is schematic only.}
\end{figure}

The distribution here is generated by 
first restricting to $|u_\parallel|\le 7\sigma_\nu$, which introduces 
negligible error since only a fraction $\sim 1.3\times 10^{-12}$ of the 
H atoms have higher velocities $|u_\parallel|$ than this.   We then use a 
rejection method with a piecewise constant comparison function.  
Specifically, we begin by defining the region in the 
$(u_\parallel,w)$-plane shown in Fig.~\ref{f:rect}.  The 
boundaries $\{u_j\}_{j=1}^6$ are chosen to enclose both the Gaussian 
(Maxwell) distribution and the \lya\ resonances: 
they are
\begin{eqnarray}
u_1 &=& -7\sigma_\nu, \nonumber \\
u_2 &=& \nu^{(1)}-\nu_E-5\gamma, \nonumber \\
u_3 &=& \nu^{(1)}-\nu_D+5\gamma, \nonumber \\
u_4 &=& \nu^{(1)}-\nu_B-5\gamma, \nonumber \\
u_5 &=& \nu^{(1)}-\nu_A+5\gamma, {\rm ~and}\nonumber \\
u_6 &=& 7\sigma_\nu.
\end{eqnarray}
if $F_i=1$.  If $F_i=0$ one substitutes $\nu_A,\nu_B\rightarrow\nu_C$ and 
$\nu_D,\nu_E\rightarrow\nu_F$, since these are the resonances that can be 
excited from $F_i=0$.

The upper limits are chosen as follows.  The non-resonant upper limit is
\begin{equation}
w_1 = \max_{j=1}^6 \phi^u_{F_iF_f}(\nu^{(1)}-u_j);
\end{equation}
this is an upper limit to 
$\rme^{-u_\parallel^2/2\sigma^2_\nu}\phi^u_{F_iF_f}(\nu^{(1)}-u)$ in the 
entire region of interest excluding the resonance regions $[u_2,u_3]$ and 
$[u_4,u_5]$, since $\rme^{-u_\parallel^2/2\sigma^2_\nu}$ never exceeds $1$ 
and $\phi^u_{F_iF_f}(\nu^{(1)}-u)$ has local maxima only at the 
resonance peaks.  The resonant upper limit is
\begin{equation}
w_2 = {\cal R} \exp \left(-\frac{1}{2\sigma_\nu^2}
|u_\parallel|_{\rm min}^2 \right),
\end{equation}
where $|u_\parallel|_{\rm min}$ is the minimum value of $u_\parallel$ in 
the resonant regions $[u_2,u_3]$ and $[u_4,u_5]$.  The amplitude ${\cal 
R}$ can be any number greater than the maximum of 
$\phi^u_{F_iF_f}(\nu')$; this guarantees that $w_2$ is an upper limit to 
$\rme^{-u_\parallel^2/2\sigma^2_\nu}\phi^u_{F_iF_f}(\nu^{(1)}-u)$ within 
the resonant regions.  Here we choose ${\cal R}$ to be $0.156/\gamma$, 
$0.078/\gamma$, $0.026/\gamma$, and $0.207/\gamma$ for $0\rightarrow 0$, 
$0\rightarrow 1$, $1\rightarrow 0$, and $1\rightarrow 1$ scattering, 
respectively.

Once the point $(u_\parallel,w)$ has been chosen, we accept it if
\begin{equation}
w<\rme^{-u_\parallel^2/2\sigma^2_\nu}\phi^u_{F_iF_f}(\nu^{(1)}-u);
\end{equation}
if this is not the case, we generate a new point.

\end{document}